\documentclass[twoside]{article}
\usepackage{qic,graphicx}

\newcommand{\be}{\begin{equation}}
\newcommand{\ee}{\end{equation}}
\newcommand{\ba}{\begin{array}}
\newcommand{\ea}{\end{array}}
\newcommand{\calA}{{\cal A }}
\newcommand{\calL}{{\cal L }}
\newcommand{\calH}{{\cal H }}

\newcommand{\calM}{{\cal M }}
\newcommand{\calN}{{\cal N }}
\newcommand{\calK}{{\cal K }}
\newcommand{\calZ}{{\cal Z }}
\newcommand{\calS}{{\cal S }}
\newcommand{\calC}{{\cal C }}
\newcommand{\la}{\langle}
\newcommand{\ra}{\rangle}

\newcommand{\LinOp}{{\bf L}}
\newcommand{\rk}[1]{\mathop{\mathrm{Rk}(#1)}}

\newtheorem{cor}{Corollary}
\newtheorem{dfn}{Definition}
\newtheorem{problem}{Problem}

\let\epsilon\varepsilon
\let\eps\varepsilon

\def\Im{\mathop{\mathrm{Im}}\nolimits}
\def\Ker{\mathop{\mathrm{Ker}}\nolimits}
\def\calA{{\mathcal A}}
\def\BB{\mathsf B}
\def\CC{\mathsf C}

\newcommand*\NP{\ensuremath{\mathrm {NP}}}
\newcommand*\QMA{\ensuremath{\mathrm {QMA}}}

\newcommand*\MA{\ensuremath{\mathrm {MA}}}

\newcommand*{\poly}{\mathop{\mathrm{poly}}}
\newcommand{\CES}{\ensuremath{\mathrm {CES}}}

\newcommand{\Square}{\begin{flushright} $\square$ \end{flushright}}

\def\sx{\sigma_x}
\def\sy{\sigma_y}
\def\sz{\sigma_z}
\begin{document}
\setlength{\textheight}{8.0truein}    

\runninghead{Commutative version of the local Hamiltonian}
            {S. Bravyi and M. Vyalyi}

\normalsize\textlineskip
\thispagestyle{empty}
\setcounter{page}{1}

\copyrightheading{0}{0}{2003}{000--000}

\vspace*{0.88truein}

\alphfootnote

\fpage{1}

\centerline{\bf
Commutative version of the local Hamiltonian problem}
\vspace*{0.035truein}
\centerline{\bf and common eigenspace problem.}
\vspace*{0.37truein}
\centerline{\footnotesize
Sergey Bravyi\footnote{e-mail: serg@cs.caltech.edu}}
\vspace*{0.015truein}
\centerline{\footnotesize\it Institute for Quantum Information, 
California Institute of Technology,}
\baselineskip=10pt
\centerline{\footnotesize\it Pasadena, CA 91125, USA.}
\vspace*{10pt}
\centerline{\footnotesize 
Mikhail Vyalyi\footnote{e-mail: vyalyi@mccme.ru}}
\vspace*{0.015truein}
\centerline{\footnotesize\it Independent University of Moscow,}
\baselineskip=10pt
\centerline{\footnotesize\it Bol'shoi Vlas'evskii per. 11, Moscow 119002, Russia}
\vspace*{0.225truein}
\publisher{(\today)}{(revised date)}

\vspace*{0.21truein}

\abstracts{
We study the complexity of a problem ``Common Eigenspace'' --- verifying
consistency of eigenvalue equations for composite quantum systems.
The input of the problem is a family of pairwise commuting Hermitian  operators
$H_1,\ldots,H_r$ on a Hilbert space $(\CC^d)^{\otimes n}$
and a string of real numbers $\lambda_1,\ldots,\lambda_r$.
The problem is to determine
whether the common eigenspace  specified by equalities
$H_a|\psi\ra=\lambda_a|\psi\ra$, $a=1,\ldots,r$ has a positive
dimension. We consider two cases: (i) all operators $H_a$
are $k$-local; (ii) all operators $H_a$ are factorized.
It can be easily shown that both problems belong to the class \QMA{} --- quantum
analogue of \NP, and that some \NP-complete problems can be reduced to
either (i) or (ii).  A non-trivial question is whether the problems (i) or (ii)
belong to \NP? We show that the answer is positive for some special
values of $k$ and $d$. Also we prove that the problem (ii) can be reduced to
its special case, such that all operators $H_a$ are factorized projectors
and all $\lambda_a=0$.}{}{}

\vspace*{10pt}

\vspace*{3pt}
\communicate{to be filled by the Editorial}

\vspace*{1pt}\textlineskip    
\section{Formulation of the problem}
\label{section_problems}        

Quantum complexity were studied intensely during the last decade. Many
quantum complexity classes were invented (to find any of them see a
comprehensive list~\cite{Aar}).
Many interesting results are known for these classes.
Nevertheless, the exact  relationship between quantum and classical complexity
classes remain open for almost all of them.
In this paper we will focus on the classical complexity class \NP{} and its
quantum analogue \QMA{} which  was defined in~\cite{KShV},~\cite{AN02}.

Let us recall the definitions
of these classes. A Boolean function $F \colon \BB^* \to \BB$ is in
\NP{} iff there is a function
$R \colon \BB^*\times\BB^* \to \BB$
computable in polynomial time on a classical
computer and a polynomial $p$ such that
$$
\ba{rcl}
F(x)=1 &\Rightarrow& R(x,y)=1 \; \mbox{for some} \; y\in \BB^*, \, |y|<p(|x|).\\
F(x)=0 &\Rightarrow& R(x,y)=0 \; \mbox{for any} \; y\in \BB^*, \, |y|<p(|x|).\\
\ea
$$
(Here and below $\BB=\{0,1\}$ and $\BB^*$ is the set of finite binary strings.
A length of string $x\in \BB^*$ is denoted by $|x|$.)
It will be convenient to
introduce two players: Arthur and Merlin. Arthur wants to compute $F(x)$,
but he is not powerful enough to do that without assistance of Merlin.
Merlin sends him the string $y$ as a `proof' that $F(x)=1$.
The properties of $R(x,y)$ guarantee that Merlin can convince Arthur that
$F(x)=1$ iff $F(x)=1$.

The class \QMA{} is defined analogously,  but Arthur is able to process quantum information.
For our purposes it suffices to mention three distinctions between \QMA{} and \NP.
Firstly, there is a quantum communication channel between Arthur and Merlin.
Thus Merlin's message may be a quantum superposition of
many strings $y$. Secondly, Arthur has a quantum computer which he uses to verify
the proof (i.e. the function $R(x,y)$ is computed by a quantum circuit, rather than 
a classical one).
Thirdly, the verification may fail with a non-zero probability.
However, the gap between Arthur's acceptance probabilities corresponding 
to $F(x)=1$ and $F(x)=0$ must be sufficiently large (bounded by a
polynomial in $1/|x|$).

By definition, $\NP \subseteq \MA \subseteq \QMA$, where \MA{} is the
class of Merlin-Arthur games --- probabilistic analogue of the
class~\NP. It is not known whether these inclusions are strict.  But
good candidates for separating $\QMA$ and $\MA$ exist. The first
example is the group non-membership problem (GNM).
Watrous~\cite{Wat00} showed that GNM in the oracle model has succint
quantum proofs. He also constructed an oracle $B$ such that
$\mathrm{GNM}(B)\notin \MA^B$. So, in a relativized world the
inclusion $\MA^B\subset \QMA^B$ is strict. The second example was
found by Aharonov and Regev~\cite{AR03}. It is a complement to a gap
version of the shortest lattice vector problem.

Similarly to the class \NP, the class \QMA{} has complete
problems. The first \QMA-complete problem was found by
Kitaev~\cite{KShV}. It is the $k$-local Hamiltonian problem with $k\geq5$.
Later Kempe and Regev~\cite{KR}
proved that the 3-local Hamiltonian problem is also \QMA-complete.
Then Kempe, Kitaev, and Regev~\cite{KKR} combined this result with a perturbative
analysis to show that the 2-local Hamiltonian is \QMA-complete.
Recently, Janzing, Wocjan and Beth have found another example
of \QMA-complete problem, see~\cite{JWB}. It is a non-identity check
for an unitary operator given by a quantum circuit.

Recall, that the input of the 2-local Hamiltonian problem
is $x=(H,\eps_l,\eps_u)$,
where $H$ is a Hermitian operator (a Hamiltonian) acting on a Hilbert space $(\CC^d)^{\otimes n}$
and $\eps_l<\eps_u$ are real numbers, such that $\eps_u-\eps_l\ge 1/\poly{(n)}$.
The operator $H$ is represented as a sum of pairwise interactions:
\begin{equation}\label{2-local}
H=\sum_{1\le a<b\le n} H_{ab}.
\end{equation}
The function $F(x)$ to be computed\footnote{Some binary encoding must be used for an input of all
problems. Accordingly, all functions to be computed are Boolean functions (may be
partially defined).} 
\ is defined as
\begin{equation}\label{LH_problem}
\ba{rcl}
F(x)=1&\Leftrightarrow& H \; \mbox{has an eigenvalue not exceeding} \; \eps_l,\\
F(x)=0&\Leftrightarrow& \mbox{all eigenvalues of} \; H \; \mbox{are greater than} \; \eps_u.\\
\ea
\end{equation}
Merlin convinces Arthur that $F(x)=1$ by sending him the ground state $|\Psi_0\ra$ of the
Hamiltonian $H$. For any Merlin's message $|\Psi\ra$ Arthur can efficiently evaluate
an expectation value $\la \Psi|H|\Psi\ra$, see~\cite{KShV}, that allows him to verify   
Merlin's proof.

For some special classes of Hamiltonians the ground state may admit a
good {\it classical } description (a good description must have a polynomial length
and must allow classical polynomial verification algorithm for Arthur).
A trivial case is a Hamiltonian $H$ such that all interactions $H_{ab}$ are
diagonal in the standard product basis of $(\CC^d)^{\otimes n}$. Then
the ground state is a basis vector. It can be described by $n\log{(d)}$
classical bits. The corresponding 2-local Hamiltonian problem thus belongs to
\NP. As an example, consider a graph $G=(V,E)$ with  qubits living at vertices
and an `antifferomagnetic'
Hamiltonian $H=+\sum_{(u,v)\in E} \sigma^z_u\sigma^z_v$, where $\sigma^z_u$
is the Pauli operator acting on the qubit $u$.
As was shown in~\cite{WB03}, it yields \NP-complete problem.
Note that generally Arthur can not solve the problem without Merlin's assistance,
because the Hamiltonian is highly frustrated.

A less restricted case of the 2-local Hamiltonian problem is obtained by
putting pairwise commutativity constraint on the individual interactions:
\begin{equation}\label{commute}
H_{ab} H_{cd} = H_{cd} H_{ab} \quad \mbox{for all pairs}\quad
(a,b) \quad \mbox{and} \quad (c,d).
\end{equation}
In this case all interactions are still diagonalized over
the same basis. In particular, the ground state $|\Psi_0\ra$ of $H$ satisfies
eigenvalue equations
\[
H_{ab}|\Psi_0\ra = \lambda_{ab}|\Psi_0\ra \quad \mbox{for all} \quad 1\le a<b\le n,
\]
while the lowest eigenvalue of $H$ is 
\[
E_0=\sum_{1\le a<b\le n} \lambda_{ab}.
\]
(If some pair of particles $a,b$ do not interact with each other, i.e.,
$H_{a,b}=0$, one can take $\lambda_{ab}=0$.)
However, a priori, there is no good classical description for the state $|\Psi_0\ra$.
Note that a list of the eigenvalues $\{\lambda_{ab}\}$ is not a good
classical description, since some configurations of the eigenvalues may be 
inconsistent due to frustrations or (and) the entanglement monogamy.
So the complexity of the problem may be higher than \NP. 

As a simple example consider Hamiltonians associated with the one-dimensional
cluster states, see~\cite{BR00}. The cluster state $|C_n\ra$ is an entangled
state of a linear chain of $n$ qubits. It is specified by eigenvalue equations
\begin{equation}\label{S_a}
S_a |C_n\ra=|C_n\ra, \quad S_a=(\sigma^z\otimes\sigma^x\otimes \sigma^z)[a-1,a,a+1],
\end{equation}
where $a$ runs from $1$ to $n$ and the square brackets indicates
the qubits acted on by an operator (we use the periodic boundary conditions
$\sigma^\alpha[0]\equiv \sigma^\alpha[n]$
and  $\sigma^\alpha[n+1]\equiv \sigma^\alpha[1]$).
All operators $S_a$ pairwise commute. Define a Hamiltonian $H$ as
\[
H=-\sum_{a=1}^n S_a.
\]
This Hamiltonian is $2$-local with respect to a coarse-grained partition,
such that the qubits $1,2$ comprise the first particle, the qubits $3,4$
--- the second, and so on (the partition is defined only for even $n$).
Its unique ground state is the cluster state $|C_n\ra$. 
This example demonstrates that the  commutativity constraint~(\ref{commute}) does
not prevent the ground state of $H$ from being highly entangled.

We shall prove that the ground state of any 2-local Hamiltonian~(\ref{2-local})
satisfying the commutativity constraint~(\ref{commute}) always admits a
good classical description\footnote{The lowest eigenvalue of $H$ may be degenerate.
In this case one can choose a ground state with a good classical description.},
so  the corresponding 2-local Hamiltonian
problem belongs to \NP{} (is \NP-complete for $d\ge 3$).
It should be contrasted with the general 2-local Hamiltonian problem,
which is \QMA-complete.

We consider here this problem and some other problems involving sets of
pairwise commuting Hermitian operators acting on a product space
\begin{equation}
\label{composite}
\calH=\calH_1\otimes\calH_2\otimes\cdots\otimes \calH_n.
\end{equation}
The factors $\calH_j$ will be referred to as `particles'.
The maximal local dimension
$$
d=\max_{j=1,\ldots, n} \dim{\calH_j}
$$
will be regarded as a constant. 
Let us introduce two classes of operators.
An operator $H\in \LinOp(\calH)$ is called {\it factorized}
if it can be expressed as $H=h_1\otimes h_2\otimes \cdots\otimes h_n$
for some $h_j\in \LinOp(\calH_j)$. 
For any group of particles $S\subseteq \{1,\ldots,n\}$ and
for any operator $h\in \LinOp(\bigotimes_{j\in S} \calH_j)$
there exists a naturally defined operator $h[S]\in \LinOp(\calH)$.
It is equal to a tensor product of $h$ with identity operators
for all $j\notin S$.
An operator $H\in \LinOp(\calH)$ is called {\it strictly $k$-local}
if it can be  expressed as 
$H=h[S]$ for some $S\subseteq\{1,\ldots,n\}$, $|S|\le k$, and
$h\in \LinOp(\bigotimes_{j\in S} \calH_j)$.
Note that if $d$ and $k$ are regarded as constants, both
factorized and $k$-local operators admit a concise classical 
description (its length grows at most linearly with $n$).

Consider now a family of Hermitian operators $H_1,\ldots,H_r\in \LinOp(\calH)$
such that
\begin{equation}\label{ab=ba}
H_a H_b=H_b H_a \quad \mbox{for all} \quad 1\le a,b\le r,
\end{equation}
and a set of real numbers $\lambda_1,\ldots,\lambda_r$.
We shall use a notation $x=(H_1,\ldots,H_r;\lambda_1,\ldots,\lambda_r)$
for all these data as it will be a typical input of our problems.
The operators $H_a$ will be referred to as {\it check operators}.
Define a {\it common eigenspace} (CES) corresponding to $x$ as
\begin{equation}\label{ces}
\calL_x = \left\{ |\psi\ra \in \calH \, : \,
H_a |\psi\ra = \lambda_a |\psi\ra \quad \mbox{for all} \quad a=1,\ldots,r\right\}
\end{equation}
If there are no vectors $|\psi\ra\in \calH$ satisfying all the eigenvalue equations,
the common eigenspace is empty, $\calL_x=0$. 

\begin{problem} \rm {\bf
(THE $k$-LOCAL CES)} \label{problem1}
The input is $x=(H_1,\ldots,H_r;\lambda_1,\ldots,\lambda_r)$, where
all check operators $H_a$ are $k$-local. 
Determine whether the
common eigenspace $\calL_x$ has a positive dimension.
\end{problem}

\begin{problem} \rm {\bf (THE FACTORIZED CES)}\label{problem2}
The input is $x=(H_1,\ldots,H_r;\lambda_1,\ldots,\lambda_r)$, where
all check operators $H_a$ are factorized. 
Determine whether the common eigenspace $\calL_x$ has a positive dimension.
\end{problem}

To analize the complexity of these problems, the input $x$ must be
represented by a binary string using a suitable encoding. 
Assuming that an eigenvalue and a matrix element of a linear operator can be represented by
a constant number of bits (see a remark at the end of this section),
the length of the input is
$|x|=O(d^{2k} r)$ for the $k$-local CES and $|x|=O(d^2 nr)$ for the factorized CES.
As was mentioned above, $d$ and $k$ are regarded as constants, so the length
of the input is bounded by a polynomial, $|x|=\poly{(n+r)}$.
Note also that the consistency of the input, i.e., the 
commutativity constraint~(\ref{ab=ba}), 
 can be verified by an algrorithm running in a time $\poly{(n+r)}$. 
If $x$ is regarded as a binary string, both problems require 
computation of a Boolean function
\begin{equation}\label{F}
\ba{rcl}
F(x)=1&\Leftrightarrow& \calL_x\ne 0,\\
F(x)=0&\Leftrightarrow& \calL_x=0.\\
\ea
\end{equation}
{\it Remarks:}
The input of the \CES{} problems consists of operators and their
eigenvalues.  Operators acting on a space of fixed dimension will be
represented by their matrix elements in some fixed basis.
Note that the \CES{} problems are formulated in terms of exact
equalities. So, we need an appropriate `exact' representation of
(complex) numbers. A good choice is algebraic numbers
of bounded degree of the extension over rationals. These numbers are
represented by arrays of rationals and we have a
trivial algorithm to check an exact equality for them.

If matrix elements are algebraic numbers and a size of the matrix is
fixed then eigenvalues of the matrix  are also algebraic numbers
(roots of a characteristic polynomial) of a bounded degree of the
extension over rationals.

To keep the bounded degree condition we put some additional
restrictions to an input of factorized CES.
Namely, we require that eigenvalues of all factors must
belong to the {\it same} extension of
bounded degree over
rational numbers. So the eigenvalues which appear
in the input belong to the same field.

It is important that such data can be efficiently manipulated.
In other words there are algorithms running in polynomial time which
solve all common linear algebra tasks in a space of bounded dimension (solving
systems of linear equations, finding eigenvalues and eigenvectors of an
operator and so on),
see books~\cite{AHU, BP94} for the subject.


\section{Summary of main results}
\label{section_results}
\noindent

Our first theorem states the upper bound on the complexity of
the \CES{} problems.  
\begin{theorem}\label{theorem_QMA}
The $k$-local and the factorized \CES{} problems belong to \QMA.
\end{theorem}
Intuitively, it follows from the fact that any state $|\psi\ra\in \calL_x$
is a sound proof that $\calL_x$ is not empty. Merlin's proving strategy
in both problems is to send Arthur an arbitrary state $|\psi\ra\in \calL_x$.
The key part of Arthur's verification algorithm is to measure eigenvalues of the
check operators, see Section~\ref{section_aux} for details.

The next theorem establishes the lower bound on the
complexity of the \CES{} problems.
\begin{theorem}\label{theorem_NP}
The $k$-local \CES{}
is   \NP-hard for
$k=2$, $d\ge 3$ or $k\ge 3$, $d\ge 2$.
The factorized \CES{} is   \NP-hard for $d\ge 2$.
\end{theorem}
We construct \NP-hard instances without resorting to quantum mechanics
at all --- the corresponding check operators are classical, that is 
diagonal in the standard product basis. Namely, we will show that 
\NP-complete problems 3-coloring and  3-CNF can be reduced to
`classical' \CES{} problems, see Section~\ref{section_aux} for details.

Our main result is that the \CES{} problems belong to \NP{} for
special values of $k$ and $d$. 

\begin{theorem}\label{theorem_main1}
The $2$-local \CES{} belongs to \NP.
\end{theorem}
We prove this theorem using the concept of interaction algebra introduced 
by Knill, Laflamme, and Viola in~\cite{KLV00} and the elementary representation
theory for finite-dimensional $\CC^*$-algebras. 
Roughly speaking, we find a fine-grained partition of each particle into
smaller subsystems which we call subparticles. These subparticles 
are naturally grouped into interacting
pairs, such that there is no interaction between different pairs. 
To verify that the common eigenspace is non zero, one suffuces to do it
for each pair of subparticles independently. It can be done efficiently. 
The fine-grained partition reveals itself only on certain subspace of $\calH$.
It can be specified locally and Merlin's proof is just a 
description of this subspace.
Amazingly, the structure of the common eigenspace resembles very much
the structure of states with ``quantum Markov chain'' property, see~\cite{HJPW04}.

It follows from Theorems~\ref{theorem_NP},\ref{theorem_main1} that the $2$-local
\CES{} is NP-complete problem for $d\ge 3$. 
Besides, Theorem~\ref{theorem_main1} has the following corollary:
\begin{cor}\label{corol0}
The problem $2$-local Hamiltonian with the 
pairwise commutativity constraint~(\ref{commute}) belongs to \NP.
\end{cor}

As far as the factorized \CES{} is concerned, we present the following results.
\begin{theorem}\label{theorem_main3}
The factorized \CES{} with $d=2$ belongs to \NP.
\end{theorem}
The proof of this theorem relies on the
explicit formula for the dimension of the common eigenspace. Although Arthur
can not use this formula to compute the dimension efficiently, sometimes
it allows him to verify that two different instances of the problem yield the common eigenspace
of the same dimension. It happens if the two instances satisfy 
simple consistency relations. We show that for any instance $x$ of the factorized \CES{}
there exist another instance $y$ consistent with $x$, such that all check operators of $y$
are diagonal in the standard product basis. Merlin's proof that $\calL_x\ne 0$
is just a description of the instance $y$ and a basis vector belonging to $\calL_y$.

To state the next theorem let us define {\it the factorized projectors \CES}.
It is the factorized \CES{} problem whose input satisfies additional constraints.

\begin{problem} \rm {\bf  (THE FACTORIZED PROJECTORS \CES)} The same as the
factorized \CES, but all check operators $H_a$ are tensor products of
orthogonal projectors and all $\lambda_a=0$.\label{problem3}
\end{problem}

We shall prove that for any factorized \CES{}
problem can be divided into two independent subproblems.
The first subproblem is the factorized \CES{} with all
check operators being tensor products of the Pauli operators $\sigma^x$, $\sigma^y$,
and $\sigma^z$. It can be solved efficiently using the stabilizer formalism,
see~\cite{NC}. The second subproblem is the factorized projectors \CES. 
Both subproblems are defined on a subspace $\calH'\subseteq \calH$. 
This subspace is defined locally and admits a good classical description.
Arthur can efficiently identify the two subproblems provided that  
Merlin sends him a description of $\calH'$. In other words, we prove that
Problem~\ref{problem2} can be non-deterministically reduced to Problem~\ref{problem3}.


\begin{theorem}\label{theorem_main2}
If the factorized projectors \CES{} with a given
$d\ge2$
belongs to \NP{}
then the factorized \CES{} with the same $d$ also  belongs  to \NP.
\end{theorem}
\noindent
We shall derive two interesting corollaries of Theorem~\ref{theorem_main2}.
\begin{cor}
\label{corrol}
The factorized \CES{} with a constraint ($\lambda_a\ne 0$ for  $1\le a\le r$)
belongs to \NP.
\end{cor}
\begin{cor}
\label{corrol1}
The factorized \CES{} with a constraint ($H_aH_b\ne 0$ for $1\le a,b\le r$) belongs to \NP.
\end{cor}

The complexity of the $k$-local and the factorized \CES{} problem for
arbitrary values of $k$ and $d$ is still unknown.
The results of Terhal and DiVincenzo on constant depth quantum circuits~\cite{TV02}
suggest that there are instances of the $k$-local \CES{} for which $\calL_x$ does not
contain a state with a good classical description.
Indeed, consider a state $|\psi\ra=U|\psi_{sep}\ra$, where
$|\psi_{sep}\ra$ is a product state and
$U$ is a quantum circuit with two-qubit gates having a depth $D$. 
If $D\ge 3$, such circuits are hard to simulate classically,
see~\cite{TV02}, so generally $|\psi\ra$  does not admit a good classical description.
Since $|\psi_{sep}\ra$ can be specified by eigenvalue equations with $1$-local check
operators,  the state $|\psi\ra$ is a one-dimensional common eigenspace for some
$2^D$-local \CES.
This argument, however, does not tell anything about the complexity of the $k$-local
\CES, since Merlin's proof need not to be a description of a state.
Some remarks on the complexity of the factorized \CES{} are made at the end of
Section~\ref{section_main2}.

The rest of the paper is organized as follows.
Section~\ref{section_aux} contains the proof of
Theorems~\ref{theorem_QMA},\ref{theorem_NP}.
Section~\ref{sec_referee} elucidates the connection between the $k$-local \CES{}
and the $k$-local Hamiltonian problems.
Theorem~\ref{theorem_main1} is proved in Section~\ref{section_main1}.
Section~\ref{section_main2} is devoted to a proof of Theorem~\ref{theorem_main2}
and its corollaries.
In Section~\ref{section_main3}
we prove that the factorized projectors \CES{}
for qubits ($d=2$) belongs to \NP. 
Being combined with Theorem~\ref{theorem_main2}, 
this result immediately implies that the factorized \CES{} for qubits
belongs to \NP, i.e., Theorem~\ref{theorem_main3}. 
Unfortunately we do not know how to generalize
the algorithm described in Section~\ref{section_main3}
to the case $d\ge 3$. 
The reason this algorithm fails for $d\ge 3$ is rather 
non-trivial and can be understood
with the help of Kochen-Specker theorem~\cite{KS}.
We briefly discuss a connection with Kochen-Specker theorem
in the concluding part of Section~\ref{section_main3}.


\section{Inclusion in \QMA{} and \NP-hardness}
\label{section_aux}
\noindent
The proof of Theorem~\ref{theorem_QMA} is contained in the
following two lemmas.
\begin{lemma} The $k$-local \CES{} belongs to \QMA.
\end{lemma}

{\bf Proof:}
Let $x=(H_1,\ldots,H_r;\lambda_1,\ldots,\lambda_r)$ be an instance of the $k$-local \CES,
$\calL_\lambda$ be the common eigenspace, and $F(x)$ be the Boolean function~(\ref{F}) 
to be computed. Merlin's proof that $F(x)=1$ will be a quantum state
$|\eta\ra\in \calH$, see~(\ref{composite}). 
We shall construct a polynomial (in $|x|$) size quantum circuit that
tells Arthur whether to accept or reject the proof (i.e. decide that 
$F(x)=1$ or $F(x)=0$). 

The Hilbert space $\calH$ can be encoded using $n\log_2{d}$ qubits.
Under this encoding any check operator $H_a$ acts non-trivially on at most
$k\log_2{d}$ qubits (this number does not depend on the complexity 
parameters $n$, $r$ and must be regarded as a constant).

One can assume without loss of generality, that all operators
$H_a$ are orthogonal projectors and all $\lambda_a=1$ (otherwise,
consider the spectral decomposition of $H_a$ and substitute $H_a$
by the projector corresponding to the eigenvalue $\lambda_a$).
Define
a POVM measurement $M_a$ corresponding to the decomposition
$I= H_a + (I-H_a)$. Since the operator $H_a$ acts only on a constant
number of qubits, Arthur can implement the measurement $M_a$ by
a quantum circuit of the size $\poly{(\log{(1/\delta)})}$, where $\delta$ is
the approximation precision, or an error probability, see~\cite{KShV}.
The parameter $\delta$ will be chosen later.
Suppose Arthur implements the measurements $M_1,\ldots, M_r$
and gets outcomes $\lambda_1',\ldots,\lambda_r'\in\{0,1\}$
(the order is not essential, since the measurements commute).
If no errors have occured, the post-measurement state $|\eta'\ra$
satisfies eigenvalue equations
$$
H_a |\eta'\ra = \lambda_a' |\eta'\ra, \quad a=1,\ldots,r.
$$
Arthur accepts the proof $|\eta\ra$ iff all $\lambda_a'=1$ (in which
case $|\eta'\ra\in \calL_\lambda$ and thus $\calL_\lambda\ne 0$). 
Note that a probability of having at least one error in the whole
verification protocol is
bounded from above by $r\delta$. The probability of the error-less
verification is thus $p_s\approx 1- r\delta$. We will choose
$\delta\ll 1/r$, so that $p_s\approx 1$.

If $F(x)=1$, Merlin can send Arthur a state $|\eta\ra\in \calL_\lambda$.
Then Arthur accepts the proof with a probability at least $p_s$.
If $F(x)=0$, Arthur may accept the proof only due to errors.
The acceptance probability in this case is at most $1-p_s$.
The size of the quantum circuit used in the protocol is bounded by $\poly{(r)}$.
It is enough to place the problem to \QMA.\Square

In the following we shall skip the details concerning the approximation precision.
In all cases considered in this paper the approximation 
precision can be easily made arbitrarily small
with only poly-logarithmic overhead.

\begin{lemma} The factorized \CES{} belongs to \QMA. \end{lemma}

{\bf Proof:}
Let $x=(H_1,\ldots,H_r;\lambda_1,\ldots,\lambda_r)$ be an instance of the factorized \CES,
$\calL_\lambda$ be the common eigenspace, $F(x)$ be the Boolean function~(\ref{F}) 
to be computed, and $|\eta\ra\in \calH$ be the
Merlin's proof that $F(x)=1$.

Arthur may pick up  $a=1,\ldots, r$ in random and check the equality
$H_a|\eta\ra=\lambda_a|\eta\ra$ for the chosen value of $a$ only.
To do that Arthur performs a destructive measurement of the
eigenvalue of $H_a$ on the state $|\eta\ra$.
If the measured eigenvalue equals $\lambda_a$, he accepts the proof,
otherwise rejects it.
Denote $p_0$ and $p_1$ probabilities for Arthur to accept the proof
provided that $F(x)=0$ and $F(x)=1$ respectively.
Let $H_a=\bigotimes_{j=1}^n H_{a,j}$. Without loss of generality we
can assume that all factors $H_{a,j}$ are Hermitian operators.
Arthur must perform $n$ separate projective eigenvalue measurements
for all factors $H_{a,j}$.
Because each factor $H_{a,j}$ acts on $\log_2{d}$ qubits,
the whole measurement can be realized by a quantum circuit of a size
$O(n)$ (recall that $d$ is regarded as a constant).
After that  Arthur computes the product of $n$ measured eigenvalues
to evaluate $\lambda_a$.

If $|\eta\ra\in \calL_\lambda$, Arthur always accepts the proof and
thus $p_1=1$. Suppose $\calL_\lambda=0$. We shall prove that
$p_0\le 1-1/r$. Let $|\eta_0\ra \in \calH$ be
the state which maximizes the acceptance probability $p_0$.
For any real vector $\chi=(\chi_1,\ldots,\chi_r)$
denote $P(\chi)\in \LinOp(\calH)$ the projector on the subspace
specified by equalities $H_a|\psi\ra=\chi_a |\psi\ra$, $a=1,\ldots,r$
(a vector $\chi$ is analogous  to an error syndrome in quantum codes
theory). The family of the projectors $P(\chi)$ defines a unity
decomposition, i.e.
$\sum_{\chi} P(\chi)=I$. Denote also
$$
a(\chi)=\la \eta_0 |P(\chi)|\eta_0\ra.
$$
For the chosen Arthur's verification algorithm we have
$$
p_0=\frac1r \sum_{a=1}^r \sum_{\chi\, : \, \chi_a =\lambda_a} |a(\chi)|^2.
$$
Changing the order of the summations we come to
$$
p_0=\frac1r\sum_\chi |a(\chi)|^2 \left( \sum_{a\, : \, \chi_a=\lambda_a} 1
\right).
$$
But since $\calL_\lambda=0$ we have $\chi_a\ne \lambda_a$ for at least one
$a=1,\ldots,r$ whenever $P(\chi)\ne 0$.
Thus
$$
p_0 \le \frac1r \sum_\chi |a(\chi)|^2 (r-1) = 1-\frac1r.
$$
So we have a gap
$p_1-p_0=1/r = \Omega (1/|x|)$ between acceptance probabilities of
positive and negative instances. As was said in the beginning
of Section~\ref{section_results}, it is enough to place the problem in
\QMA. \Square

The following two lemmas constitute a proof of Theorem~\ref{theorem_NP}.
\begin{lemma}
The $2$-local \CES{} is \NP-hard for
$d\ge 3$. 
\end{lemma}
{\bf Proof:}
We will show that the \NP-complete
3-coloring problem can be reduced to $2$-local \CES{} with $d=3$.
(An idea used in this reduction was suggested by P. Wocjan in~\cite{WB03}).
Let $G=(V,E)$ be an arbitrary graph. The 3-coloring problem is to
determine whether the graph $G$ admits a coloring of the vertices
with 3 colors such that each edge has endpoints of different colors.
Let $n=|V|$  and $r=3|E|$.
Choose a Hilbert space $\calH=(\CC^3)^{\otimes n}$ such that each vertex
of the graph carries a  space $\CC^3$.
The operators $H_a$ will be assigned to the edges
with three operators assigned to each edge. These operators are
responsible for three forbidden coloring of the edge. It is convenient to
introduce a composite index $a=(uv,c)$, where $(uv)\in E$
is an edge and $c\in \{1,2,3\}$ is a color. Then the $2$-local \CES{}
$(H_1,\ldots,H_r;\lambda_1,\ldots,\lambda_r)$ is defined as
\begin{equation}\label{coloring-proj}
H_{uv,c}=(|c,c\ra\la c,c|)[u,v], \quad
\lambda_{uv,c}=0, \quad (uv)\in E, \quad c=1,2,3.
\end{equation}
Obviously, existence of non-trivial common eigenspace $\calL_\lambda$
is equivalent to existence of 3-coloring for the graph $G$.
(Note that the projectors~(\ref{coloring-proj}) also provide an instance of
the factorized projectors \CES{}.) We have shown that
$2$-local \CES{} with $d\ge 3$ is \NP-hard.\Square

\begin{lemma}
The $k$-local \CES{}
is   \NP-hard for
$d=2$, $k\ge 3$.
\end{lemma}
{\bf Proof:}
We will prove that \NP-complete 3-CNF problem can be reduced
to $3$-local \CES{} with $d=2$. Recall that 3-CNF
(conjunctive normal form)  is a Boolean function
of the form $L(x)=C_1(x)\wedge C_2(x)\wedge\cdots \wedge C_r(x)$,
$x=(x_1,\ldots,x_n)\in \BB^n$,
where each {\it clause } $C_a(x)$ is a disjunction of three
literals (a literal is a variable or negation of a variable).
An example of three-literal clause is $x_1\vee x_3\vee (\neg x_5)$.
The 3-CNF problem is to determine whether an equation $L(x)=1$
admits at least one solution.
Choose a Hilbert space
$\calH=(\CC^2)^{\otimes n}$. The operators $H_a$ and the
eigenvalues $\lambda_a$ must be assigned to
the clauses $C_a(x)$ according to the following table:

\begin{center}
\begin{tabular}{|c|c|c|}
\hline
$C_a(x)$ & $H_a$ & $\lambda_a$ \\
\hline
$x_i\vee x_j\vee x_k$ & $(|0,0,0\ra\la 0,0,0|)[i,j,k]$& 0\\
$x_i\vee x_j\vee (\neg x_k)$ & $(|0,0,1\ra\la 0,0,1|)[i,j,k]$& 0\\
$\cdots$ & $\cdots$ & $\cdots$ \\
$(\neg x_i)\vee(\neg x_j)\vee(\neg x_k)$ & $(|1,1,1\ra\la 1,1,1|)[i,j,k]$& 0\\
\hline
\end{tabular}
\end{center}

\noindent
It is easy to check that the common eigensubspace
for the $3$-local \CES{} introduced above is non-trivial
iff the equation $L(x)=1$ has at least one solution.
Thus we have reduced 3-CNF problem to the $3$-local \CES.\Square

Obviously, the $3$-local \CES{} assigned to 3-CNF problem in the
previous lemma is a special case of the factorized projectors \CES{}
(and thus a special case of the factorized \CES). So we have proved all
statements of  Theorem~\ref{theorem_NP}.

\section{The $k$-local commuting Hamiltonian}
\label{sec_referee}

We shall now discuss the $k$-local Hamiltonian problem. Recall that
the problem is to evaluate the Boolean function~(\ref{LH_problem})
with the Hamiltonian
\begin{equation}\label{k_local}
H=\sum_{a=1}^r H_a, \quad
H_a \; \mbox{is strictly {\it k}-local for all } a.
\end{equation}
If, additionally, all terms in $H$ pairwise commute,
$$
H_a H_b = H_b H_a \quad \mbox{for all} \quad  a,b,
$$
we shall call the problem ``$k$-local commuting Hamiltonian''.
The goal of this section is to reduce the $k$-local commuting Hamiltonian
to the $k'$-local \CES. In the first Lemma a non-determistic reduction
with $k'=k$ is put forward. It also shows that Corollary~\ref{corol0}
indeed follows from Theorem~\ref{theorem_main1}. 
The second Lemma~\cite{referee} establishes a deterministic reduction with $k'=k+1$.

\begin{lemma}\label{lemma:5}
If the $k$-local \CES{} belongs to \NP{} then
the $k$-local commuting Hamiltonian also belongs to \NP.
\end{lemma}
{\bf Proof:}
Obviously, we can choose a complete set of eigenvectors of $H$ which are eigenvectors
of all operators $H_a$ also. To prove that $H$ indeed has an eigenvalue
not exceeding $\eps_l$ Merlin can send Arthur a set of eigenvalues
$(\lambda_1,\ldots,\lambda_r)$ such that

\noindent
(i) $\sum_{a=1}^r \lambda_a \le \eps_l$,

\noindent
(ii) $(H_1,\ldots,H_r;\lambda_1,\ldots,\lambda_r)$ is a positive instance
of $k$-local \CES{} (i.e. $\calL_\lambda\ne 0$).

\noindent
Although Arthur can not verify (ii) by himself, according to assumption of the
lemma this verification belongs to \NP. So Arthur can ask Merlin
to include a proof of (ii) in his message.
It follows that $k$-local commuting Hamiltonian problem belongs to \NP.\Square

\begin{lemma}\label{referee}
The problem $k$-local commuting Hamiltonian can be polynomially
reduced to the $(k+1)$-local \CES.
\end{lemma}
{\bf Proof:}
Let $x=(H,\eps_l,\eps_u)$ be an instance of 
the $k$-local commuting Hamiltonian. Here the Hamiltonian $H$
has the form~(\ref{k_local}).
Taking the spectral decomposition of each operator $H_a$
we can rewrite the Hamiltonian as follows:
\[
H=\sum_{a=1}^R \eps_a \Pi_a, \quad
\Pi_a \Pi_b = \Pi_b \Pi_a \quad \mbox{for all} \quad  a,b,
\]
where all $\Pi_a$ are orthogonal projectors. Note that the number of terms $R$
is at most $R=rd^k$, that is only linear in the length of the input $|x|$
(recall that $d$ and $k$ 
are regarded as constants). For any binary string $y=(y_1,\ldots,y_R)$
define the corresponding energy
\[
E(y)=\sum_{a=1}^R \eps_a y_a,
\]
and the eigenspace
\[
\calL_y = \{ |\psi\ra\in \calH \, : \,
\Pi_a|\psi\ra = y_a |\psi\ra \quad \mbox{for all} \quad a=1,\ldots,R\}.
\]
Then $x$ is a positive instance of the problem iff
there exist a binary string $y$ such that $E(y)\le \eps_l$ and
$\calL_y\ne 0$. Let us define a partially defined Boolean function
\begin{equation}
\ba{rcl}
R(y)=1&\Leftrightarrow& E(y)\le \eps_l, \\
R(y)=0&\Leftrightarrow& E(y) > \eps_u.\\
\ea
\end{equation}
Obviously, $R(y)$ can be computed by an algorithm running in 
a polynomial time, or equivalently, there exists a polynomial classical circuit 
that computes $R(y)$. It allows to cast the function $R(y)$ into a 3-CNF
with  only a polynomial number of clauses:
\begin{equation}\label{R(y)}
R(y)={C_1(y)}\wedge {C_2(y)}\wedge \cdots \wedge {C_M(y)}, \quad M=\poly{(|x|)}.
\end{equation}
Here each clause $C_j$ involves at most three bits $y_a$.
(For a connection between classical circuits and 3-CNFs see~\cite{KShV}.)
We are now ready to present an instance of the $(k+1)$-local \CES{} 
associated with $x$. The \CES{} problem is defined on the space
\[
\calH'=\calH\otimes (\CC^2)^{\otimes R}.
\]
The auxiliary $R$ qubits will `keep' the binary string $y$. 
Denote
$|0_a\ra\la 0_a|$ and $|1_a\ra\la 1_a|$ the projectors $|0\ra\la 0|$
and $|1\ra\la 1|$ applied to the $a$-th qubit.
The \CES{} problem has two families of check operators. The first one is
\[
H_a'=\Pi_a\otimes |1_a\ra\la 1_a| + (I-\Pi_a)\otimes |0_a\ra\la 0_a|,
\quad a=1,\ldots R.
\]
Roughly speaking, $H_a'$ ties the value of $y_a$ to the eigenvalue of the projector $\Pi_a$.
Note that the operators $H_a'$ are strictly $(k+1)$-local.
The check operators of the second family act only on the qubits. They
are associated with the clauses $C_j$ in~(\ref{R(y)}). Let us
introduce an operator $\hat{C}_j$ acting on $R$ qubits such that
its action on the basis vectors $|y\ra\in (\CC^2)^{\otimes R}$ is
\[
\hat{C}_j |y\ra=C_j(y)|y\ra.
\]
The corresonding check operator acting on $\calH'$ is
$I\otimes \hat{C}_j$. It is strictly $3$-local.
Consider a common eigenspace
\[
\calM = \{|\psi\ra\in \calH' \, : \,
H_a'|\psi\ra=|\psi\ra, \quad I\otimes \hat{\calC}_j |\psi\ra=|\psi\ra
\quad \mbox{for all} \quad a=1,\ldots, R; \; 
j=1,\ldots, M\}.
\]
It follows from the definitions that $\calM\ne 0$ iff 
there exist a product state $|\psi\ra\otimes |y\ra\in \calH'$
such that $|\psi\ra\in \calL_y$ and $R(y)=1$. It means that
$x$ is a positive
instance of the $k$-local commuting Hamiltonian problem.\Square


\section{The 2-local common eigenspace problem}
\label{section_main1}
\noindent

Let us start from revisiting the  example of 
cluster states, see Section~\ref{section_problems}.
Recall that the chain of $n$ qubits is partitioned into two-qubit particles as shown on Fig.~1.
There are $n$ check operators $S_1,\ldots, S_n$, see~(\ref{S_a}).
The common eigenspace $\calL$ is defined by equations $S_a|\psi\ra=|\psi\ra$, where
$a$ runs from $1$ to $n$. In this example $\calL$ is one-dimensional with
the basis vector $|C_n\ra$. Although $|C_n\ra$ is a highly entangled state, its
entanglement has very simple structure with respect to the coarse-grained partition.
Indeed, denote the qubits comprising the $j$-th particle as
$j.l$ and $j.r$, see Fig.~1. A pair of qubits $j.r$ and $(j+1).l$ will be refered to as a bond.
Let $V_j$ be the controlled-$\sigma^z$
operator applied to the qubits $j.l$ and $j.r$, and $V=V_1\otimes\cdots \otimes V_n$.
It is an easy exercise to verify that the state $V|C_n\ra$ is a tensor product 
over the bonds:
\begin{equation}\label{cluster_state}
V|C_n\ra = |\phi[1.r,2.l]\ra\otimes |\phi[2.r,3.l]\ra\otimes\cdots
\otimes |\phi[n.l,1,r]\ra,
\end{equation}
where the square brackets indicate owners of a state and $|\phi\ra\in \CC^2\otimes \CC^2$
is specified
by eigenvalue equations $(\sigma^x\otimes\sigma^z)|\phi\ra=
(\sigma^z\otimes\sigma^x)|\phi\ra=|\phi\ra$. 
In other words, $|C_n\ra$ can be prepared from a collection of bipartite
pure states distributed between the particles by local unitary operators.
This fact is not just a coincidence.
We will show later that for any instance of the $2$-local \CES{} 
the common eigenspace is either empty or 
contains a state  which can be created from a collection of
bipartite pure states by applying local
isometries (local unitary embeddings into a larger Hilbert space).

\begin{figure}[b]
\centerline{\epsfig{file=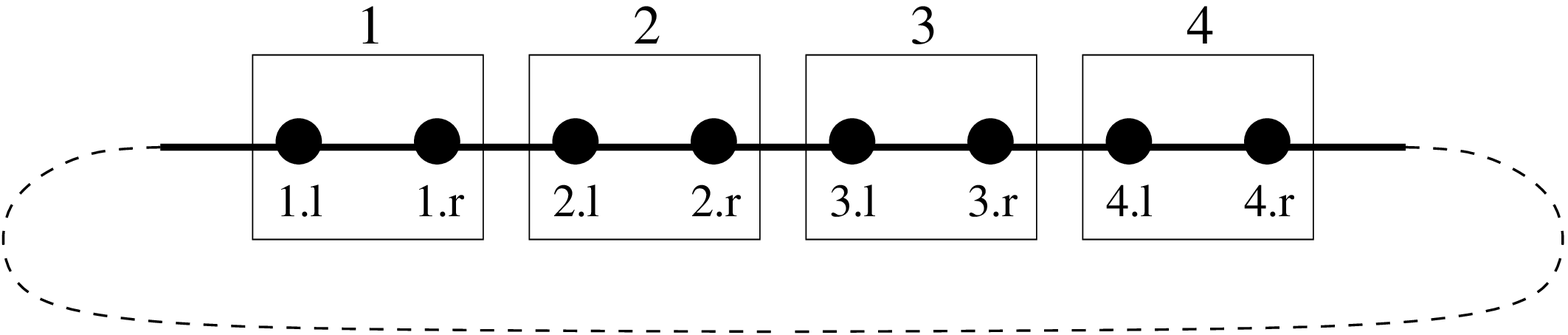, width=8.2cm}} 
\vspace*{13pt}
\fcaption{\label{Figure1}A chain of $8$ qubits is partitioned into
$n=4$ particles with local dimensions $d=4$.}
\end{figure}

We continue by making three simplifications that allow one to
reduce the number of check  operators.
Let $x=(H_1,\ldots,H_r;\lambda_1,\ldots,\lambda_r)$ be an instance of the $2$-local \CES
and $\calL_x$ be the common eigenspace.

{\it Simplification~1:} Clearly, $\calL_x=0$ unless $\lambda_a$ is an eigenvalue of 
$H_a$. Since Arthur can verify it efficiently, we shall assume that
the input of the $2$-local \CES{} satisfies an additional
constraint: 
\[
\lambda_a\in \mbox{Spec}(H_a) \quad \mbox{for all} \quad a=1,\ldots, r.
\]

{\it Simplification~2:} It eliminates all check operators acting only on one
particle. Suppose that the check operator $H_a$ acts only on the particle $j$ i.e.,
$H_a=h[a]$ for some $h\in \LinOp(\calH_j)$. The eigenvalue equation 
$H_a|\psi\ra=\lambda_a|\psi\ra$ implies that the space $\calH_j$ can be reduced
 to the eigenspace $\Ker{(h-\lambda_a I)}\subseteq \calH_j$. 
Indeed, denote 
\[
\calH_l'=\left\{ \begin{array}{rcl} \calH_l &\mbox{for}& l\ne j, \\
                                    \Ker{(h-\lambda_a I)} &\mbox{for} & l=j,\\
\end{array}\right.
\quad \mbox{and} \quad
\calH'=\bigotimes_{j=1}^n \calH_j'\subseteq \calH.
\]
It is clear that $\calL_x\subseteq \calH'$. Moreover, since all check operators
commute, the subspace $\calH'$ is preserved by all of them, so one can define
the restrictions
\[
H_b'=H_b|_{\calH'} \in \LinOp(\calH'), \quad b=1,\ldots,r.
\]
Since the reduction $\calH\to \calH'$ is done locally, all operators $H_b'$
are strictly $2$-local. Also, they all pairwise commute. It may happen
however that $\lambda_b\notin \mbox{Spec}(H_b')$ for some $b$. 
If this is the case, one has $\calL_x=0$. Otherwise, we arrive to a new instance
of the $2$-local \CES{} $y=(H_1',\ldots,H_r';\lambda_1,\ldots,\lambda_r)$ 
which is equivalent to $x$. Since $H_a'=\lambda_a I$, the corresponding  eigenvalue
equation is trivial and the pair $(H_a';\lambda_a)$ can be excluded from $y$.
We have reduced the number of check operators by one and the dimension of
some particle at least by one. Obviously, Arthur can implement this reduction efficiently.
After at most $r$ iterations Arthur either decides that $\calL_x=0$ or arrives to
a simplified instance in which all check operators act non-trivially on
two particles.

{\it Simplification~3:}
We will show now that all operators $H_a$ acting on some particular pair of
particles $(j,k)$ can be substituted by a single check operator.
Indeed, let us group the operators $H_1,\ldots, H_r$
into subsets $S_{jk}$, $1\le j<k\le n$, such that
$S_{jk}$ contains all labels $a$ for which $H_a$ acts on the particles $j$
and $k$. To distinguish the pairs for which $S_{jk}\ne \emptyset $ we shall
characterize an instance of the $2$-local \CES{} by its {\it interaction graph}
$G=(V,E)$, such that $V$ is the set of particles, and
edges are drawn between interacting particles.
\begin{dfn}
A graph $G=(V,E)$ with $V=\{1,2,\ldots, n\}$ and 
$E=\{ (j,k) \, : \, S_{jk}\ne \emptyset\}$ is called an interaction graph of
the instance $x$.
\end{dfn}
For any  $(j,k)\in E$ consider an eigenspace 
\[
\calL_{jk}=\{ |\psi\ra\in \calH \, : \, H_a |\psi\ra = \lambda_a |\psi\ra
\quad \mbox{for all} \quad a\in S_{jk}\}.
\]
Denote $\Pi_{jk}\in \calL(\calH)$ the orthogonal projector onto $\calL_{jk}$.
Clearly, $\{\Pi_{jk}\}_{(j,k)\in E}$ is a family of pairwise commuting $2$-local operators and
the common eigenspace $\calL_x$ can be specified by  equations 
\begin{equation}\label{L_x}
\calL_x=\{ |\psi\ra\in \calH \, : \, \Pi_{jk}|\psi\ra=|\psi\ra \quad \mbox{for all}
\quad (j,k)\in E\}.
\end{equation}
Thus $x$ is equivalent to an instance
\begin{equation}\label{Pi_{jk}}
y=(\{\Pi_{jk}\}_{(j,k)\in E};1,\ldots,1).
\end{equation}
Summarizing the three simplifications above, one suffices to prove Theorem~\ref{theorem_main1}
only for the following version of the $2$-local \CES.

\noindent
{\bf Input:} An interaction graph $G=(V,E)$  and
a family of $2$-local pairwise commuting projectors $x=\{\Pi_{jk}\}_{(j,k)\in E}$.
For every pair $(j,k)\in E$  the projector 
$\Pi_{jk}$ acts non-trivially on both $\calH_j$ and $\calH_k$
(in particular $\Pi_{jk}\ne 0$).

\noindent
{\bf Problem:} Determine whether the common eigenspace~(\ref{L_x})
has a positive dimension.

Our first goal is to introduce a notion of {\it irreducible} instance and prove
Theorem~\ref{theorem_main1} for irreducible instances only. Then 
we will generalize the proof to arbitrary instances.
\begin{dfn}\label{dfn_irr}
Let $x=\{\Pi_{jk}\}_{(j,k)\in E}$  be an instance
of the $2$-local \CES. Consider a subalgebra 
$\calN_j\subseteq\LinOp(\calH_j)$ of operators acting on the particle $j$
and commuting with all check operators:
\begin{equation}\label{inv}
\calN_j = \{ O\in \LinOp(\calH_j) \, : \, 
O[j] \Pi_{jk} = \Pi_{jk} O[j] \quad \mbox{for all} \quad (j,k)\in E\}.
\end{equation}
The instance $x$ is called irreducible iff algebras $\calN_j$ are 
trivial i.e., $\calN_j=\CC\cdot I$ for all $j=1,\ldots,n$. 
\end{dfn}
{\it Remark: }   Arthur can check whether an instance is irreducible 
using an efficient algorithm (the constraints~(\ref{inv}) are
given by linear equations on a space of bounded dimension).
We shall now prove that  any irreducible
instance of the $2$-local \CES{} is positive ($\calL_x\ne 0$).
The proof is based on the following lemma.

\begin{lemma}\label{lemma:2CES}
Let $x=\{\Pi_{jk}\}_{(j,k)\in E}$ be an irreducible instance of the $2$-local \CES{}
with an interaction graph $G=(V,E)$. There exist 
\begin{itemize}
\item A pair of Hilbert spaces $\calH_{j.k}$ and $\calH_{k.j}$ 
associated with each edge $(j,k)\in E$,
\item A tensor product structure $\calH_j = \bigotimes_{k\, : (j,k)\in E} \calH_{j.k}$,
\end{itemize}
such that the projector $\Pi_{jk}$ acts non-trivially only on the two factors
$\calH_{j.k}\otimes \calH_{k.j}$ in the decomposition 
$\calH = \bigotimes_{l=1}^n \bigotimes_{m\, : \, (l,m)\in E}\calH_{l.m}$.
\end{lemma}

\noindent
The lemma says that there exist a fine-grained partition of the system,
such that the particle $j$ is decomposed into several subparticles
$\{j.k\}$, where $(j,k)\in E$. The interaction between the particles $j$ and $k$
affects only the subparticles $j.k$ and $k.j$, that is 
$\Pi_{jk}=h_{jk}[j.k,k.j]$ for some $h_{jk}\in \LinOp(\calH_{j.k}\otimes \calH_{k.j})$.
A straightforward corollary of the lemma is that 
the common eigenspace $\calL_x$ has a tensor product structure:
\begin{equation}\label{ps1}
\calL_x = \bigotimes_{(j,k)\in E} \calM_{jk}, 
\end{equation}
where $\calM_{jk}\subseteq \calH_{j.k}\otimes \calH_{k.j}$ is specified by an
equation $h_{jk} |\psi\ra=|\psi\ra$. 
Since $\Pi_{jk}\ne 0$ for $(j,k)\in E$, one has $h_{jk}\ne 0$, and thus
$\calM_{jk}\ne 0$, which implies $\calL_x \ne 0$. 
So the lemma has the following amazing corollary.
\begin{cor}
Any irreducible instance of the $2$-local \CES{} is
positive.
\end{cor}

Now we move on to the proof of Lemma~\ref{lemma:2CES}.
The main mathematical tool used in the analysis is the representation theory for
finite-dimensional $\CC^*$-algebras. In the subsequent discussion the term 
$\CC^*$-algebra refers to any algebra of operators on a finite-dimensional Hilbert
space  which is $\dag$-closed and contains the identity.
The center of a $\CC^*$-algebra $\calA$ will be denoted $Z(A)$. By definition,
\[
Z(A)=\{X\in \calA\, : \, XY=YX \quad \mbox{for all} \quad Y\in \calA\}.
\]
An algebra has a trivial center iff $Z(A)=\CC\cdot I$. 
We shall use the following fact (for the proof see the book~\cite{Takesaki},
or Theorem~5 in~\cite{KLV00}):

\noindent
{\bf Fact~1:}  {\it 
Let $\calH$ be a Hilbert space and $\calA\subseteq \LinOp(\calH)$ be
a $\CC^*$-algebra with a trivial center. There exists a tensor
product structure $\calH=\calH_1\otimes\calH_2$
such that $\calA$ is the subalgebra of all operators acting on the
factor $\calH_1$ i.e.,
$$
\calA= \LinOp(\calH_1)\otimes I.
$$
}

\noindent
{\bf Proof of Lemma~\ref{lemma:2CES}:}
Consider any pair $(j,k)\in E$ and let
$\Pi_{jk}=h[j,k]$ for some $h\in \LinOp(\calH_j\otimes\calH_k)$, $h\ne 0$. 
Our goal is to construct two $\CC^*$-algebras $\calA_{j.k}\subseteq \LinOp(\calH_j)$
and  $\calA_{k.j}\subseteq \LinOp(\calH_k)$ such that $h\in \calA_{j.k}\otimes \calA_{k.j}$.
The main element of the construction was proposed by Knill, Laflamme, and 
Viola~\cite{KLV00}, who studied $\dag$-closed algebras generated by an interaction between
a system and an environment.
Consider a decomposition
\begin{equation}\label{H=sum}
h=\sum_{\alpha} A_\alpha \otimes B_\alpha,
\end{equation}
where the families of operators $\{A_\alpha \in \LinOp(\calH_j)\}$
and $\{B_\alpha \in \LinOp(\calH_k)\}$  are linearly independent.
Denote $\calM_{j.k}$ and $\calM_{k.j}$ the linear spaces spanned by
$\{A_\alpha\}$ and $\{B_\alpha\}$ respectively. One can easily verify that
$\calM_{j.k}$ and $\calM_{k.j}$ do not depend upon the choice of
the decomposition~(\ref{H=sum}). An identity
\[
h^\dag = h = \sum_{\alpha} A_\alpha^\dag \otimes B_\alpha^\dag,
\]
tells us that $\calM_{j.k}$ and $\calM_{k.j}$ are
closed under Hermitian conjugation.
Define $\calA_{j.k}\subseteq \LinOp(\calH_j)$ and  $\calA_{k.j}\subseteq \LinOp(\calH_k)$
as the minimal $\CC^*$-algebras such that $\calM_{j.k}\subseteq\calA_{j.k}$
and $\calM_{k.j}\subseteq\calA_{k.j}$. Equivalently,  $\calA_{j.k}$ is
generated by the family $\{A_\alpha\}\cup I$ and
$\calA_{k.j}$ is generated by $\{B_\alpha\}\cup I$.
(The fact that $h$ is a projector is irrelevant for this construction.)

Consider any triple of particles $j\ne k \ne l$ such that 
$(j,k)\in E$ and $(j,l)\in E$. What can be said about the $\CC^*$-algebras
$\calA_{j.k},\calA_{j.l}\subseteq\LinOp(\calH_j)$?

The first claim is that these algebras commute i.e.,
\begin{equation}\label{XY=YX}
XY=YX \quad \mbox{for all} \quad X\in \calA_{j.k} \quad \mbox{and}
\quad Y\in \calA_{j.l}.
\end{equation}
Indeed, the projectors $\Pi_{jk}$ and $\Pi_{jl}$ can be represented as
\[
\Pi_{jk}=H[j,k,l],\quad  \Pi_{jl}=G[j,k,l],
\]
where the operators $H,G\in \LinOp(\calH_j\otimes\calH_k\otimes\calH_l)$ admit
decompositions
\[
H=\sum_\alpha A_\alpha\otimes B_\alpha \otimes I, \quad
G=\sum_\beta C_\beta \otimes I \otimes D_\beta.
\]
Here all the families $\{A_\alpha\}$, $\{B_\alpha\}$, $\{C_\beta\}$, and
$\{D_\beta\}$ are linearly independent. The commutativity constraint
$\Pi_{jk}\Pi_{jl}=\Pi_{jl}\Pi_{jk}$ yields
\[
\sum_{\alpha,\beta} (A_\alpha C_\beta - C_\beta A_\alpha) \otimes B_\alpha \otimes D_\beta =0.
\]
All terms in the sum are linearly independent due to the second and the third factors.
Thus the equality is possible only if $A_\alpha C_\beta = C_\beta A_\alpha$ for
all $\alpha$ and $\beta$. Since the algebras $\calA_{j.k}$ and $\calA_{j.l}$ are generated by
$\{A_\alpha\}$ and $\{C_\beta\}$ respectively, we conclude that they commute.

The next step is to prove that the center $Z(\calA_{j.k})$
is trivial for all $(j,k)\in E$. Indeed, it follows from~(\ref{XY=YX})
that any central element $Z\in Z(\calA_{j.k})$ commutes with all elements of the algebras
$\calA_{j.l}$, where $(j,l)\in E$. Since $\Pi_{jl}=h[j,l]$ for some
$h\in \calA_{j.l}\otimes \calA_{l.j}$, we conclude that an operator
$Z[j]\in \LinOp(\calH)$ commutes with all projectors $\Pi_{jl}$.
Since we consider an irreducible
instance of \CES, it is possible only if $Z=\lambda\cdot I$ for some
complex number $\lambda$. Thus $Z(\calA_{j.k})=\CC\cdot I$.

Let us show how $\calH_j$ acquires the tensor product structure for
some particular $j$. For any pair $(j,k)\in E$ one can make use of Fact~1
with $\calH\equiv \calH_j$ and $\calA\equiv \calA_{j.k}\subseteq \LinOp(\calH_j)$.
It follows that $\calH_j$ admits a decomposition
\begin{equation}\label{decomp'}
\calH_j=\calH_{j.k}\otimes \calH_j', 
\end{equation}
such that the algebra $\calA_{j.k}$ is the algebra of all operators acting on the
factor $\calH_{j.k}$ i.e.,
\begin{equation}\label{decomp''}
\calA_{j.k}=\LinOp(\calH_{j.k})\otimes I.
\end{equation}
Consider now a third particle $l$ such that $(j,l)\in E$. Let us examine the
commutativity relation between the algebras $\calA_{j.k}$ and $\calA_{j.l}$.  
It is consistent with the decompositions~(\ref{decomp'},\ref{decomp''}) iff
$\calA_{j.l}$ acts trivially on the factor $\calH_{j.k}$.
In other words, any element $X\in \calA_{j.l}$ has a form $X=I\otimes X'$
for some $X'\in \LinOp(\calH_j')$. We can now make use of Fact~1 with
$\calH\equiv \calH_j'$ and $\calA\equiv \calA_{j.l}$ to get a finer
decomposition
\[
\calH_j=\calH_{j.k}\otimes \calH_{j.l}\otimes \calH_j'',
\]
such that
\[
\calA_{j.k}=\LinOp(\calH_{j.k})\otimes I \otimes I \quad \mbox{and} \quad
\calA_{j.l} = I\otimes \LinOp(\calH_{j.l}) \otimes I.
\]
Repeating these arguments we arrive to a decomposition
$\calH_j=(\bigotimes_{k\, : \, (j,k)\in E} \calH_{j.k})\otimes \calH_{j.j}$,
such that the algebra $\calA_{j.k}$ coincides with the algebra of all
linear operators on the factor $\calH_{j.k}$. As for the last factor $\calH_{j.j}$,
it is acted on by neither of the algebras. This factor however can not appear for
an irreducible problem. Indeed, any operator $X\in \LinOp(\calH_j)$ acting only
on $\calH_{j.j}$ would commute with all algebras $\calA_{j.k}$. Accordingly, an operator
$X[j]$ would commute with all projectors $\Pi_{jk}$. This is possible only if
$X=\lambda\cdot I$. Thus the algebra $\LinOp(\calH_{j.j})$ is just the algebra of
complex numbers. It follows that $\calH_{j.j}=\CC$ and it can be removed from the
decomposition. Summarizing, we get
\[
\calH_j=\bigotimes_{k\, : \, (j,k)\in E} \calH_{j.k}, \quad
\calA_{j.k}=I\otimes \cdots \otimes I\otimes \LinOp(\calH_{j.k})\otimes I
\otimes \cdots \otimes I.
\]
It follows from the
definitions above that $\Pi_{jk}$ acts non-trivially only on the
factor $\calH_{j.k}$ in $\calH_j$ and only on the factor $\calH_{k.j}$ in 
$\calH_k$. The lemma is proved.
\Square

The next step is to generalize Lemma~\ref{lemma:2CES} to reducible instances.
We first outline the generalization and then put it formally.
For each particle $j$ 
a local `classical variable' $\alpha_j$ will be defined.  Each value of $\alpha_j$
specifies a subspace $\calH_j^{\alpha_j}\subseteq \calH_j$, such that
a decomposition $\calH_j=\bigoplus_{\alpha_j}\calH_j^{\alpha_j}$ is a direct sum.
This decomposition is preserved by all check operators. If one fixes the classical
variables $\alpha_1,\ldots,\alpha_n$ for each particle, one gets some
subspace $\calH^{(\alpha_1\ldots\alpha_n)}\subseteq\calH$. The restriction of the
problem on this subspace is almost irreducible (in the sense specified below),
so Lemma~\ref{lemma:2CES} can be applied. In other words,
for fixed values of the classical variables the fine-grained partition into subparticles
emerges. The subparticles are naturally grouped into pairs, such that there is no any
interactions between different pairs. Arthur can solve the restricted problem efficiently.
Accordingly, the role of Merlin is just to send
Arthur the values of the classical variables $\alpha_1,\ldots,\alpha_n$ for which
the intersection $\calL_x \bigcap \calH^{(\alpha_1 \ldots \alpha_n)}$ is not empty.
\begin{lemma}\label{lemma:2CESfull}
Let $x=\{\Pi_{jk}\}_{(j,k)\in E}$ be an instance of the $2$-local \CES{}
with an interaction graph $G=(V,E)$. There exist 
\begin{itemize}
\item Direct sum decompositions $\calH_j=\bigoplus_{\alpha_j} \calH_j^{(\alpha_j)}$ with
induced decomposition $\calH=\bigoplus_\alpha \calH^{(\alpha)}$, where
$\alpha\equiv(\alpha_1,\ldots,\alpha_n)$ and
$\calH^{(\alpha)}=\calH_1^{(\alpha_1)}\otimes\cdots\otimes \calH_n^{(\alpha_n)}$,
\item A pair of Hilbert spaces $\calH_{j.k}^{(\alpha_j\alpha_k)}$ and 
$\calH_{k.j}^{(\alpha_k\alpha_j)}$ associated with each edge $(j,k)\in E$,
\item Hilbert spaces $\calH_{j.j}^{(\alpha_j)}$,
\item A tensor product structure $\calH_j^{(\alpha_j)} =\calH_{j.j}^{(\alpha_j)}\otimes
\left( \bigotimes_{k\, : (j,k)\in E} 
\calH_{j.k}^{(\alpha_j\alpha_k)}\right)$,
\end{itemize}
such that the check operators admit a decomposition
\[
\Pi_{jk}=\bigoplus_{\alpha} \Pi_{jk}^{(\alpha_j\alpha_k)},
\]
where $\Pi_{jk}^{(\alpha_j\alpha_k)}\in \LinOp(\calH^{(\alpha)})$ acts only on the factors
$\calH_{j.k}^{(\alpha_j\alpha_k)}\otimes \calH_{k.j}^{(\alpha_k\alpha_j)}$ in the
tensor product
\begin{equation}\label{structure_of_H_j}
\calH^{(\alpha)}=\left( \bigotimes_{l=1}^n \calH_{l.l}^{(\alpha_l)} \right) \otimes
\left(\bigotimes_{l=1}^n \, \, 
 \bigotimes_{m\, : \, (l,m)\in E} \calH_{l.m}^{(\alpha_l\alpha_m)}\right).
\end{equation}
\end{lemma}

As in Lemma~\ref{lemma:2CES}, the notation $j.k$ refers to subparticles of the
particle $j$. 
It should be noted that the spaces $\calH_{j.j}^{(\alpha_j)}$ are acted on by 
neither of the check operators. That is why they do not appear in Lemma~\ref{lemma:2CES}. 
However, if the problem is reducible, and there exist
an operator $h[j]$ commuting with all check operators, it acts {\it only} on the
spaces $\calH_{j.j}^{(\alpha_j)}$.
Also it should be mentioned that any of the Hilbert spaces listed in Lemma~\ref{lemma:2CESfull}
may be one-dimensional.

A straightforward corollary of the lemma is that the common eigenspace can be
represented as a direct sum:
\begin{equation}\label{ps2'}
\calL_x=\bigoplus_{\alpha}
\calM^{(\alpha)},\quad
\calM^{(\alpha)} = \calL_x\bigcap \calH^{(\alpha)}
\end{equation}
where each subspace $\calM^{(\alpha)}$ has a tensor product structure:
\begin{equation}\label{ps2''}
\calM^{(\alpha)}=\left( \bigotimes_{j=1}^n \calH_{j.j}^{(\alpha_j)}\right)\otimes
\left( \bigotimes_{(j,k)\in E} \calM_{jk}^{(\alpha_j\alpha_k)}\right),
\quad 
\calM_{jk}^{(\alpha_j\alpha_k)}\subseteq \calH_{j.k}^{(\alpha_j\alpha_k)}\otimes
\calH_{k.j}^{(\alpha_k\alpha_j)}.
\end{equation}
(Some of the subspaces $\calM_{jk}^{(\alpha_j\alpha_k)}$ may be zero though.)
Indeed, the lemma says that 
$\Pi_{jk}^{(\alpha_j\alpha_k)}=h_{jk}^{(\alpha_j\alpha_k)}[j.k,k.j]$ for some 
$h_{jk}^{(\alpha_j\alpha_k)}\in \LinOp\left(\calH_{j.k}^{(\alpha_j\alpha_k)}\otimes 
\calH_{k.j}^{(\alpha_k\alpha_j)}\right)$.
Thus the eigenvalue equations $\Pi_{jk}|\psi\ra=|\psi\ra$ specifying $\calL_x$
lead to~(\ref{ps2'},\ref{ps2''}) with 
\begin{equation}\label{ps2'''}
\calM_{jk}^{(\alpha_j\alpha_k)} =\left\{ |\phi\ra\in \calH_{j.k}^{(\alpha_j\alpha_k)}\otimes
\calH_{k.j}^{(\alpha_k\alpha_j)} \, : \, 
h_{jk}^{(\alpha_j\alpha_k)}|\phi\ra=|\phi\ra \right\}.
\end{equation}

Theorem~\ref{theorem_main1} is a simple corollary of Lemma~\ref{lemma:2CESfull}.
Indeed, Merlin's proof that $\calL_x\ne 0$ may be a description of the subspaces
$\calH_j^{(\alpha_j)}\subseteq \calH_j$,
$j=1,\ldots,n$, such that $\calL_x\bigcap \calH^{(\alpha)}\ne 0$.
Arthur uses Merlin's message to find the restricted projectors
$\Pi_{jk}^{(\alpha_j\alpha_k)}$. It follows from~(\ref{ps2'},\ref{ps2''},\ref{ps2'''}) that
$\calL_x\ne 0$ iff $\Pi_{jk}^{(\alpha_j\alpha_k)}\ne 0$ for all $j$ and $k$.
Arthur can verify it efficiently. 

Besides, Lemma~\ref{lemma:2CESfull} implies that the common eigenspace $\calL_x$
contains a state with a good classical description. Indeed, choose some value
of $\alpha$ for which $\calL_x\bigcap \calH^{(\alpha)}\ne \emptyset$. Denote
$V_j\, : \, \calH_j^{(\alpha_j)}\to \calH_j$ an isometry corresponding to the
embedding $\calH_j^{(\alpha_j)}\subseteq \calH_j$.
Choose an arbitrary state $|\phi_{jk}\ra\in \calM_{jk}^{(\alpha_j\alpha_k)}$ and
an arbitrary state $|\phi_j\ra\in \calH_{j.j}^{(\alpha_j)}$. 
Denote
\[
|\phi\ra=\bigotimes_{j=1}^n |\phi_j\ra \otimes \left(
\bigotimes_{(j,k)\in E} |\phi_{jk}\ra \right) \in \calH^{(\alpha)}.
\]
This state is just a collection of bipartite pure states and local unentangled states.
As such it has a concise classical description.
A state $|\phi'\ra= (V_1\otimes\cdots\otimes V_n)|\phi\ra$
belongs to $\calL_x$ and also has a concise classical description. 
An eigenvalue equation $\Pi_{jk}|\phi'\ra=|\phi'\ra$ follows from identities
\[
\Pi_{jk} V  = V \Pi_{jk}^{(\alpha_j\alpha_k)}, \quad
\Pi_{jk}^{(\alpha_j\alpha_k)}|\phi\ra=|\phi\ra,
\]
where we denoted $V=V_1\otimes \cdots \otimes V_n$.

In the rest of this section we prove Lemma~\ref{lemma:2CESfull}.
It requires  a generalization of Fact~1 to $\CC^*$-algebras with non-trivial center
(the statement given below coincides with Theorem~5 in~\cite{KLV00}).

\noindent
{\bf Fact~2:} {\it
Let $\calH$ be a Hilbert space and $\calA\subseteq\LinOp(\calH)$ be a $\CC^*$-algebra.
There exist a direct sum decomposition
$\calH=\bigoplus_{\alpha} \calH^{(\alpha)}$ and a tensor product structure
$\calH^{(\alpha)}=\calH_1^{(\alpha)}\otimes\calH_2^{(\alpha)}$ such that 
\[
\calA=\bigoplus_\alpha \LinOp(\calH_1^{(\alpha)})\otimes I.
\]
The center $Z(\calA)$ is generated by orthogonal projectors on 
the subspaces $\calH^{(\alpha)}$. }

\noindent
{\bf Proof of Lemma~\ref{lemma:2CESfull}:}
Define $\CC^*$-algebras $\calA_{j.k}\subseteq \LinOp(\calH_j)$ for $(j,k)\in E$
in the same way as in the proof of Lemma~\ref{lemma:2CES}. The key role is played by a
$\CC^*$-algebra $\calA_{j.j}\equiv \calN_j\subseteq \LinOp(\calH_j)$,
see~(\ref{inv}). These algebras obey certain commutativity relations.
Namely, 
\begin{equation}\label{XY=YX'}
XY=YX \quad \mbox{for all} \quad X\in \calA_{j.k} \quad \mbox{and} \quad
Y\in \calA_{j.l},
\end{equation}
whenever $j\ne k\ne l$, $(j,k)\in E$, $(j,l)\in E$, or 
$j=k\ne l$, $(j,l)\in E$. They follow  either from~(\ref{XY=YX}) or from
the definitions. It follows that any element of the center $Z(\calA_{j.k})$ commutes
with all algebras under consideration. As such, it must be an element of
$\calA_{j.j}$, that is $Z(\calA_{j.k})\subseteq \calA_{j.j}$. But the algebras $\calA_{j.k}$
and $\calA_{j.j}$ pairwise commute, so one has
\begin{equation}\label{ZZ}
Z(\calA_{j.k})\subseteq Z(\calA_{j.j}) \quad \mbox{for all} \quad
(j,k)\in E.
\end{equation}
Let us apply Fact~2 with $\calA\equiv \calA_{j.j}$ and $\calH\equiv \calH_j$.
One gets a direct sum decomposition
\begin{equation}\label{H=HK}
\calH_j = \bigoplus_{\alpha_j} \calH_j^{(\alpha_j)}, \quad
\calH_j^{(\alpha_j)} =\calH_{j.j}^{(\alpha_j)}\otimes \calK_j^{(\alpha_j)},
\end{equation}
such that
\begin{equation}\label{A^alpha}
\calA_{j.j}= \bigoplus_{\alpha_j} \LinOp(\calH_{j.j}^{(\alpha_j)})\otimes I \equiv
\bigoplus_{\alpha_j} \calA_{j.j}^{(\alpha_j)}.
\end{equation}
Consider now an edge $(j,k)\in E$. It follows from~(\ref{XY=YX'}) that any element of
$\calA_{j.k}$ preserves the subspaces $\calH_j^{(\alpha_j)}$. Thus the algebra
$\calA_{j.k}$ has the same direct sum structure:
\[
\calA_{j.k}=\bigoplus_{\alpha_j} \calA_{j.k}^{(\alpha_j)}, \quad
\calA_{j.k}^{(\alpha_j)}\subseteq \LinOp(\calH_j^{(\alpha_j)}).
\]
It follows from~(\ref{ZZ}) that each subalgebra $\calA_{j.k}^{(\alpha_j)}$ has
a trivial center. Moreover, the commutativity relation~(\ref{XY=YX'}) implies that
$\calA_{j.k}^{(\alpha_j)}$ acts only on the factor $\calK_j^{(\alpha_j)}$ in the
decomposition~(\ref{H=HK}).

Let us fix any $\alpha=(\alpha_1,\ldots,\alpha_n)$ and consider
a subspace $\calH^{(\alpha)}=\bigotimes_{j=1}^n \calH_j^{(\alpha_j)}\subseteq \calH$.
Since the check operator $\Pi_{jk}$ is generated by the algebras $\calA_{j.k}$
and $\calA_{k.j}$ (see the proof of Lemma~\ref{lemma:2CES}), 
the decomposition $\calH=\bigoplus_\alpha \calH^{(\alpha)}$ is preserved by all
check operators. Therefore one can define restricted check operators
\begin{equation}
\label{Pi^alpha}
\Pi_{jk}^{(\alpha_j\alpha_k)}=\Pi_{jk}|_{\calH^{(\alpha)}}\in \LinOp(\calH^{(\alpha)}).
\end{equation}
From~(\ref{H=HK}) one gets
\begin{equation}\label{H=HK'}
\calH^{(\alpha)}=\left( \bigotimes_{j=1}^n \calH_{j.j}^{(\alpha_j)}\right)
 \otimes \calK^{(\alpha)},
\quad \calK^{(\alpha)}\equiv 
\bigotimes_{j=1}^n \calK_j^{(\alpha_j)}.
\end{equation}
It follows that the restricted check operators~(\ref{Pi^alpha})
act only on the factor $\calK^{(\alpha)}$. 

Consider an instance $y$ of the $2$-local \CES{} with the Hilbert space $\calK^{(\alpha)}$ and
the check operators~(\ref{Pi^alpha}). This instance is irreducible. Indeed, suppose
an operator $Z\in\LinOp(\calK_j^{(\alpha_j)})$
belongs to the set $\calN_j$ (see Definition~\ref{dfn_irr}) for the instance $y$.
Denote $Z'=I\otimes Z\in \LinOp(\calH^{(\alpha)})$, where $I$ acts on the first $n$ factors 
$\calH_{j.j}^{(\alpha_j)}$ in the decomposition~(\ref{H=HK'}). 
By definition, $Z'\in \calA_{j.j}^{(\alpha_j)}$, see~(\ref{A^alpha}).
But we know that the algebra $\calA_{j.j}^{(\alpha)}$ acts only on the factor
$\calH_{j.j}^{(\alpha_j)}$ in the decomposition~(\ref{H=HK'}). Thus $Z$ is proportional to the
identity, that is $y$ is irreducible.
Applying Lemma~\ref{lemma:2CES} to $y$ we get the desired decomposition~(\ref{structure_of_H_j}).
\Square

\section{The factorized common eigenspace problem}
\label{section_main2}
\noindent
In this section we prove Theorem~\ref{theorem_main2}. First of all we
shall answer a simple question: under what circumstances do factorized Hermitian
operators commute with each other?
\begin{lemma}\label{lemma_ab=ba}
Let $H_1, H_2\in \LinOp(\calH)$ be tensor products of Hermitian operators:
$$
H_a=\bigotimes_{j=1}^n H_{a,j}, \quad H_{a,j}^\dag=H_{a,j}, \quad a=1,2, \quad j=1,\ldots,n.
$$
Then the commutator $[H_1,H_2]=0$ iff one of the following conditions hold
\begin{enumerate}
\item $H_{1,j}H_{2,j}=\pm H_{2,j}H_{1,j}$ for each $j$ in the range $1,\ldots,n$.
The number of anticommuting factors is even.

\item $H_{1,j}H_{2,j}=0$ for some $j\in [1,n]$. Equivalently, $H_1 H_2 =0$.
\end{enumerate}
\end{lemma}
{\bf Proof:}
Obviously, either of conditions stated in the lemma is sufficient.
 Suppose that $[H_1,H_2]=0$ and prove that
at least one of the conditions is true. We have
\begin {equation}
\label{t-product}
\bigotimes_{j=1}^n H_{1,j}H_{2,j} = \bigotimes_{j=1}^n H_{2,j}H_{1,j}.
\end{equation}
If both sides of this equality equal zero then $H_{1,j}H_{2,j}=0$ for at least
one $j\in [1,n]$. Suppose that both sides are non-zero operators, i.e.
$H_{1,j}H_{2,j}\ne 0$ for all $j$.
Then by definition of a tensor product,
there exists a set of complex numbers $r_1,\ldots,r_n$ such that
\begin {equation}
\label{r_j}
H_{1,j}H_{2,j}=r_j H_{2,j}H_{1,j}, \quad j=1,\ldots,n \quad \mbox{and} \quad
\prod_{j=1}^n r_j =1.
\end{equation}
This equality says that the operator $H_{2,j}$ maps any eigenvector
of $H_{1,j}$ to an eigenvector of $H_{1,j}$. Under this map an eigenvalue of $H_{1,j}$ is
multiplied by $r_j$. It means that
$r_j$ must be a real number. Taking Hermitian conjugation of~(\ref{r_j}) we get
an equality $H_{2,j}H_{1,j}=r_j H_{1,j}H_{2,j}$. Combining it with~(\ref{r_j})
yields $r_j^2=1$, i.e. $r_j=\pm 1$, which completes the proof.
\Square

This lemma motivates the following definition.
\begin{dfn}\rm
Let $H_1,H_2\in \LinOp(\calH)$ be Hermitian factorized commuting operators.
We say that $H_1$ and $H_2$ commute {\it in  a singular way} iff
$H_1H_2=0$. Otherwise we say that $H_1$ and $H_2$ commute {\it in a
regular way}.
\end{dfn}
\noindent Thus saying that $H_1$ and $H_2$ commute in a regular way
implies that all factors of $H_1$ and $H_2$ either commutes or
anticommutes.

Let $x=(H_1,\ldots,H_r;\lambda_1,\ldots,\lambda_r)$ be an instance of
the factorized \CES{} problem. By definition,
\begin {equation}
\label{H_a}
H_a = \bigotimes_{j=1}^n H_{a,j}, \quad H_{a,j}^\dag = H_{a,j} \quad\mbox{for all}
\quad a=1,\ldots,n, \quad j=1,\ldots,n.
\end{equation}
It will be convenient to define a table $T_x=\{H_{a,j}\}$
whose entries are Hermitian operators.
Let us agree that the columns of the table $T_x$ correspond
to particles (the index $j$), while the rows correspond to the check operators
(the index $a$).
Let us give one more definition:
\begin{dfn}\rm
A row $a$ of the table $T_x$ is called {\it regular} if $\lambda_a\ne 0$.
If $\lambda_a=0$ the row $a$ is called {\it singular}.
\end{dfn}
\noindent
Generally, some rows of $T_x$ commute in a regular way and
some rows commute in a singular way.
Note that two regular rows always commute in a regular way unless
$\calL_x=0$. Indeed, if $H_aH_b=0$ for some regular rows $a,b$, then
for any $|\psi\ra\in \calL_x$ one has 
$0=H_a H_b |\psi\ra= \lambda_a\lambda_b |\psi\ra$. Since $\lambda_a,\lambda_b\ne 0$,
this is possible only if $|\psi\ra=0$. 
It is the presence of rows which commute in a singular way
which makes the problem highly non-trivial. In this case
the operators $H_{a,j}$ and $H_{b,j}$ may neither commute nor
anticommute and their eigenspaces may be embedded into $\calH_j$
more or less arbitrarily. In this situation we can not expect that
the common eigenspace $\calL_x$ contains a state which has a `good' classical
description.

As before, Merlin claims that $x$ is a positive instance 
($\calL_x\ne 0$) and Arthur must verify it.
First of all we note that Arthur may perform two significant simplifications
of the table $T_x$ by himself.

\noindent
{\it Simplification 1:}
Note that
$\Im{H_a} = \bigotimes_{j=1}^n \Im{H_{a,j}}$ for any $a\in [1,r]$
and that the subspace $\Im{H_a}$ is preserved by all other check operators.
If the $a$-th row is a regular one then, in addition,
$\calL_x \subseteq \Im{H_a}$. Thus we can restrict the problem on the
subspace $\calH'\subseteq \calH$ defined as
\begin {equation}
\calH'=\bigcap_{a \, : \, \lambda_a \ne 0} \Im{H_a} =
\bigotimes_{j=1}^n \calH_j', \quad
\calH_j'=
\bigcap_{a \, : \, \lambda_a \ne 0} \Im{H_{a,j}}.
\end{equation}
Obviously, restricted  check operators $H_a|_{\calH'}$
are factorized and pairwise commuting. Thus the modified problem is
the factorized \CES{} with a constraint that {\it an
operator $H_{a,j}$ is non-degenerated whenever $a$ is a regular row}.
Since Arthur can easily find the subspaces $\calH_j'$ and the restricted
operators $H_a|_{\calH'}$, we can assume that the original instance
$x$ already satisfies this constraint.

\noindent
{\it Simplification 2:}  For any singular row $b$ denote
$H_{b,j}'\in \LinOp(\calH_j)$ a projector on the subspace
$\Im{H_{b,j}}\subseteq\calH_j$. Denote
$$
H_b' = \bigotimes_{j=1}^n H_{b,j}'.
$$
Obviously, $\Im{H_b}=\Im{H_b'}=\bigotimes_{j=1}^n \Im{H_{b,j}}$, so that
\begin{equation}
\label{H2proj1}
\Ker{H_b}=\Ker{H_b'}.
\end{equation}
The subspace $\Im{H_b'}$ is preserved by all check operators $H_a$, so that
\begin{equation}
\label{H2proj2}
[H_a,H_b']=0 \quad \mbox{for all } a=1,\ldots,r.
\end{equation}
Thus if we substitute each $H_{b,j}$ by $H_{b,j}'$ (i.e.
substitute $H_b$ by $H_b'$), the new family of operators is
pairwise commuting. So it corresponds to some factorized \CES{}
problem. The equality~(\ref{H2proj1}) tells us that
both problems have the same answer.
Applying, if necessary, the substitutions $H_b\to H_b'$,
we can assume that the original problem $x$ satisfies
the following constraint: {\it $H_{b,j}$ is a projector
whenever $b$ is a singular row.}
In other words, we can assume that singular rows of the table $T_x$
constitute a factorized projectors \CES.

\begin{lemma}
\label{reg-sing}
If $a$ is a regular row and $b$ is a singular row then
$[H_{a,j},H_{b,j}]=0$ for all $j=1,\ldots,n$.
\end{lemma}
{\bf Proof:}
Since the operators $\{H_{a,j}\}_j$ are non-degenerated, we have $H_a H_b\ne 0$, i.e.
a regular and  a singular row can commute only in a regular way.
Thus $H_{a,j}$ and $H_{b,j}$ either commute or anticommute for all $j$.
Suppose that $H_{a,j}H_{b,j}=-H_{b,j}H_{a,j}$ for some $j$.
Since $H_{a,j} H_{b,j}\ne 0$, the operator $H_{a,j}$ maps an eigenvector
of $H_{b,j}$ to an eigenvector of $H_{b,j}$ reversing a sign of the eigenvalue.
But after the simplifications $H_{b,j}$ became a projector and
thus it can not anticommute with $H_{a,j}$.
\Square

Let us summarize the results of the two simplifications:
\begin{itemize}
\item $H_{a,j}$ is non-degenerated whenever $a$ is a regular row.

\item $H_{a,j}$ is a projector whenever $a$ is a singular row.

\item $[H_{a,j},H_{b,j}]=0$ for all $j$ whenever $a$ is regular and $b$ is singular.
\end{itemize}

In the remaining part of the section we describe a non-deterministic reduction
of the simplified factorized \CES{} problem to the factorized projectors \CES.
The reduction is based on the following possible transformations of the
table $T_\Lambda$ and the vector $\{\lambda_a\}$:

\begin{enumerate}
\item[\sf (i).]
 Suppose there exists $j\in [1,n]$ and a Hermitian operator
$Z\in \LinOp(\calH_j)$ such that $Z$ commutes with all
$H_{1,j},\ldots,H_{r,j}$. Then $Z[j]$ commutes with all $H_1,\ldots,H_r$ and thus
preserves the subspace $\calL_x$.
Assuming that $\calL_x\ne 0$, the operator $Z$ has some eigenvalue $\omega$
such that the intersection $\calL_x\bigcap \Ker{(Z[j]-\omega)}$
is non-zero. So a transformation
$$
\calH_j\to \calH_j'\equiv\Ker{(Z-\omega I)}\quad \mbox{and} \quad
H_{a,j}\to H_{a,j}|_{\calH_j'}, \quad a=1,\ldots,r
$$
leads to an equivalent instance. To implement this transformation,
Merlin should send a description of $(j,Z,\omega)$ to Arthur.

\item[\sf (ii).]  Suppose  for some $j\in [1,n]$ we have $\calH_j=\calH_j'\otimes\calH_j''$
and $H_{a,j}=H_{a,j}'\otimes H_{a,j}''$ for all $a=1,\ldots,r$
(here $H_{a,j}'$ acts on the factor $\calH_j'$ and
$H_{a,j}''$ acts on the factor $\calH_j''$). A transformation
replacing the $j$-th column by two new columns with entries
$\{H_{a,j}'\}$ and $\{H_{a,j}''\}$ leads to an equivalent problem.

\item[\sf (iii).] Suppose  in some column $j$ all operators $H_{a,j}$ are
proportional to the identity: $H_{a,j}=r_a I$
for some real numbers $r_a$, $a=1,\ldots,r$.
We may delete the $j$-th column from the table and perform a
transformation $\lambda_a\to \lambda_a/r_a$, $a=1,\ldots,r$.

\item[\sf (iv).] For any column $j$ we can perform a transformation
$$
H_{a,j}\to UH_{a,j}U^\dag,\quad a=1,\ldots,r,
$$
where $U\in \LinOp(\calH_j)$ is an arbitrary unitary operator.

\item[\sf (v).] For any non-zero real number $r$ we can replace some $H_{a,j}$ by
$rH_{a,j}$ and replace $\lambda_a$ by $r\lambda_a$.

\item[\sf (vi).] Swaps of the columns and swaps of the rows.

\end{enumerate}
We claim that the transformations $\sf (i)-(vi)$ allow to
transform the simplified instance $x$ into a
{\it canonical} form $x_c$.
The instance $x_c$ consists
of two independent problems. The first problem is
the factorized \CES{} with $\lambda_a=\pm 1$ and all check operators being tensor products of
the Pauli operators and the identity. 
The second problem is the factorized projectors \CES.
More strictly, the table $T_{x_c}$ for the instance $x_c$ has the following structure:

\vspace{3mm}
\begin {center}
\begin{tabular}{|p{2cm}|p{3cm}|p{17mm}}
\cline{1-2}
\begin {center} Pauli operators \end{center}&
\begin{center} $I$ \end{center} &
\begin {center}  $\lambda_a=\pm 1$ \end{center}\\
\cline{1-2}
\begin {center} $I$ \end{center}&
\begin {center}  factorized projectors  \end{center}&
\begin {center} $\lambda_a=0$ \end{center}\\
\cline{1-2}
\end{tabular}
\end{center}

\vspace{3mm}
\noindent
The table is divided into four blocks.
Columns in the left half of the table represent the qubits, i.e. $\calH_j=\CC^2$.
All operators $H_{a,j}$ sitting at the north-west block
are either the Pauli operators $\sx,\sy,\sz$, or the identity.
All operators $H_{a,j}$ sitting at the south-east block are
projectors. Any operator $H_{a,j}$ sitting in the blocks labeled by `$I$'
is the identity. The whole Hilbert space $\calH$ factorizes:
$\calH=\calH'\otimes \calH''$, where the factor $\calH'=\CC^2\otimes \cdots\otimes
\CC^2$ corresponds to the left half and $\calH''$ --- to the right half of the table.
The common eigenspace also factorizes: $\calL_{x_c}=\calL'\otimes \calL''$, where
$\calL'$ is a code subspaces of some stabilizer code (see~\cite{KShV,NC} for the subject),
 and $\calL''$ is
the factorized projectors \CES. 
Obviously $\calL_{x_c}\ne 0$ iff $\calL'\ne 0$ and $\calL''\ne 0$.
Arthur can verify that $\calL'\ne 0$ (and even compute the dimension of $\calL'$)
using an efficient algorithm, see~\cite{KShV}.
Thus the original instance $x$ has been reduced to an instance of the factorized
projectors \CES. Summarizing, Theorem~\ref{theorem_main2} follows from the claim given above.
We restate it here as a lemma.

\begin{lemma}\label{lemma_canonical}
The transformations $\sf (i)-(vi)$ allow one to
transform any instance of the factorized \CES{} 
into the canonical form.
\end{lemma}
{\bf Proof:}
Let $T_x$ be a table representing a simplified instance of the factorized \CES.
The first step is to apply the transformation~$\sf (i)$ as long as it is possible.
To describe operators $Z$ suitable for
the transformation~$\sf (i)$
it is convenient to use a language of $\CC^*$-algebras.
\begin{dfn}
A column algebra $\calA_j\subseteq\LinOp(\calH_j)$ of a column $j$ is the
$\CC^*$-algebra generated by the operators $H_{a,j}$
for all regular rows $a$.
\end{dfn}
\noindent
Let $\calZ(\calA_j)\subseteq \calA_j$ be a center of the column algebra $\calA_j$.
By definition, any operator $Z\in \calZ(\calA_j)$ commutes with all $H_{a,j}$ for regular $a$.
On the other hand, $Z$ commutes with all $H_{b,j}$ for singular $b$, see
Lemma~\ref{reg-sing}. Thus Arthur can use {\it any} operator $Z\in \calZ(\calA_j)$ to
implement the transformation~$\sf (i)$. 
We would like to choose $Z$ such that after the transformation~$\sf (i)$ the column
algebra of the column $j$ would have a trivial center.
Making use of Fact~2 from Section~\ref{section_main1}
one can identify a direct sum decompositions $\calH_j=\bigoplus_\alpha \calH_j^{(\alpha)}$
such that $\calA_j=\bigoplus_\alpha \calA_j^{(\alpha)}$, where the algebra
$\calA_j^{(\alpha)}\subseteq \LinOp(\calH_j^{(\alpha)})$ has a trivial center.
Let us apply transformation~$\sf (i)$, where  $Z$ is the projector onto $\calH_j^{(\alpha)}$
($\alpha$ can be chosen arbitrarily) and $\omega=1$.
The column algebra of the column $j$ for the transformed problem is
obviously $\calA_j^{(\alpha)}$. It has a trivial center.
Arthur must implement $n$ transformations~$\sf (i)$ for all columns $j$.
Now we can assume that all column algebras $\calA_j$ have a trivial center~\footnote{
Since Arthur can find the direct sum decompositions of $\calH_j$ and $\calA_j$
efficiently (recall that the space $\calH_j$ has a bounded dimension),
Merlin can just tell him what of the subspaces $\calH_j^{(\alpha)}$ has to be chosen.}.

Then according to Fact~1 from Section~\ref{section_main1}, the spaces $\calH_j$ have
a tensor product structure
\begin {equation}
\label{tensor}
\calH_j=\calH_j'\otimes\calH_j'',
\end{equation}
such that the column algebra $\calA_j$ acts on the factor $\calH_j'$ only:
$$
\calA_j = \LinOp(\calH_j')\otimes I.
$$
Take some singular row $b$. The operator $H_{b,j}$ commutes with all
elements of $\calA_j$, see Lemma~\ref{reg-sing}.
It means that $H_{b,j}$ acts only on the factor $\calH_j''$:
$$
H_{b,j}=I\otimes H_{b,j}''\quad \mbox{whenever } \lambda_b=0,
$$
for some operator $H_{b,j}''\in\LinOp(\calH_j'')$.
Since $H_{b,j}$ is a projector, the same does $H_{b,j}''$.
Summarizing, the whole space $\calH$ has a tensor product structure
$$
\calH=\calH'\otimes\calH'',\quad
\calH'=\bigotimes_{j=1}^n \calH_j',\quad
\calH''=\bigotimes_{j=1}^n \calH_j'',
$$
such that all regular rows act only on $\calH'$ while all singular rows
act only on $\calH''$. Applying $\poly{(n+r)}$ transformations~$\sf (ii)$, $\sf (iii)$,
and~$\sf (vi)$ we can split the 
original instance $x$ into 
two independent instances: $x'$ (regular rows) and $x''$ (singular rows),
such that $\calL_x=\calL_{x'}\otimes\calL_{x''}$.
One remains to prove that $x'$ is equivalent to
non-triviality check for some stabilizer quantum code.

Since we have already known that all singular rows can be isolated,
let us assume that all rows of the table $T_x$ are regular.
Thus all operators $H_{a,j}$ are non-degenerated and all column
algebras $\calA_j$ have a trivial center. Applying, if necessary,
the transformation~$\sf (iii)$ we can get rid of `free' factors $\calH_j''$
in~(\ref{tensor}), so we can also assume that
$$
\calA_j=\LinOp(\calH_j).
$$
For any column $j$ the operators $H_{a,j}$ either commute or anticommute
with each other. It follows that the operator $H_{a,j}^2$ belongs to the center of $\calA_j$.
Thus $H_{a,j}^2\sim I$. Applying, if necessary,
the transformation~$\sf (v)$ we can make $H_{a,j}^2=I$ for all $a$ and $j$.
Note that $\lambda_a=\pm 1$ for all $a$ after this transformation, otherwise $\calL_x=0$
by obvious reasons. A connection with stabilizer  codes is established by the
following lemma (we shall prove it later):

\begin{lemma}
\label{lemma_Pauli}
Let $\calS$ be a Hilbert space, $G_1,\ldots,G_r\in \LinOp(\calS)$ be
Hermitian operators such that
$$
G_a^2=I, \quad G_aG_b=\pm G_bG_a\quad \mbox{for all } a,b,
$$
and such that the algebra generated by $G_1,\ldots,G_r$ coincides with
$\LinOp(\calS)$.
Then there exists an integer $n$, a tensor product structure
$\calS=(\CC^2)^{\otimes n}$ and
a unitary operator $U\in \LinOp(\calS)$ such that
$UG_aU^\dag$ is a tensor product of the Pauli operators and the identity 
(up to a sign) for all $a$.
\end{lemma}

Take $\calS=\calH_j$ and $G_a=H_{a,j}$  for some column $j$.
Let $U\in \LinOp(\calH_j)$ be a unitary operator whose existence is
guaranteed by Lemma~\ref{lemma_Pauli}. Applying the transformations~$\sf (iv)$
with the operator $U$ followed by the transformation~$\sf (ii)$ to
the $j$-th column we split it into $n$ columns. Each of new columns
represents a qubit. The entries of all new columns are either the Pauli
operators or the identity. Performing this transformation for all
columns independently, we transform the original instance of the factorized \CES{} to
the factorized \CES{} with all check operators being tensor products 
of the identity and the Pauli operators.
The total number of transformations~$\sf (i)-(vi)$ that we made is $\poly{(n+r)}$.
\Square

{\bf Proof of Lemma~\ref{lemma_Pauli}:}
The family $G_1,\ldots,G_r$ contains
at least one anticommuting pair $G_aG_b=-G_bG_a$, since
otherwise the algebra generated by $G_a$'s has a non-trivial center.
Without loss of generality, $G_1G_2=-G_2G_1$. The operator $G_1$ has
only eigenvalues $\pm 1$ and $G_2$ swaps the subspaces corresponding
to the eigenvalue $+1$ and $-1$. Thus both subspaces have the same dimension
and we can introduce a tensor product  structure $\calS=\CC^2\otimes \calS'$ such that
$$
UG_1U^\dag = \sz\otimes I, \quad UG_2U^\dag =\sx\otimes I,
$$
for some unitary operator $U\in \LinOp(\calS)$.
Using the fact that all other $G_a$'s either commute or anticommute with
$G_1$ and $G_2$ one can easily show that each $G_a$ also has a product form:
$$
UG_aU^\dag=\tilde{G}_a\otimes G_a', \quad \tilde{G}_a\in\{I,\sx,\sy,\sz\},\quad
G_a'\in \LinOp(\calS').
$$
Obviously, the family of  operators $G_1',\ldots,G_r'$ satisfies
\begin {equation}
\label{induct1}
(G_a')^\dag=G_a',\quad (G_a')^2=I,\quad G_a'G_b'=\pm G_b'G_a'.
\end{equation}
Denote $\calA\subseteq\LinOp(\calS')$ the $\CC^*$-algebra generated
by the operators $G_1',\ldots,G_r'$. It has a trivial center.
Indeed, if $Z\in \calA$ is a non-trivial central element then $I\otimes Z$ is
a non-trivial central element of $\LinOp(\calS)$, which is impossible.
Applying Fact~1 from Section~\ref{section_main1} to the pair $(\calS',\calA)$,
we conclude that there exists a tensor product structure
$$
\calS'=\calS''\otimes\calS''', \quad \calA=\LinOp(\calS'')\otimes I.
$$
But the factor $\calS'''$ is acted on by neither of $G_a$'s
and thus $\calS'''=\CC$. We have proved that
\begin {equation}
\label{induct2}
\calA=\LinOp(\calS').
\end{equation}
Taking into account~(\ref{induct1}) and~(\ref{induct2}) we
can apply induction with respect to $\dim{\calS}$
(the base of induction corresponds to $\calS=\CC$).
\Square

We conclude this section by proving Corollaries~\ref{corrol} and~\ref{corrol1}.
Obviously, if $\lambda_a\ne 0$ for all $a$ then all rows of  the table $T_x$
are regular and thus the factorized \CES{} can be non-deterministically
reduced to
non-triviality check for an additive quantum code.
Suppose now that $H_aH_b\ne 0$ for all $a$ and $b$. It means that all rows
of the table (both regular and singular) commute in a regular way.
Thus the factorized
projectors \CES{} which appears in our reduction has the following special
property: for any column $j$ all projectors $H_{a,j}$ pairwise commute.
Therefore the space $\calH_j$ has a basis in which all projectors $H_{a,j}$
are diagonal. So the problem becomes classical and belongs to \NP{}
by obvious reasons.

\section{The factorized projectors common eigenspace problem for qubits}
\label{section_main3}
\noindent
In this section  we prove that the factorized projectors \CES{} for qubits
($d=2$)  belongs to \NP. 
 Let us start from a general note that applies to an arbitrary $d$.
Consider an instance $x=(H_1,\ldots,H_r)=\{H_{a,j}\}$ of the factorized projectors \CES{} and the
common eigenspace
\[
\calL_x=\{|\psi\ra\in \calH\, : \, H_a|\psi\ra=0 \quad \mbox{for all} \quad a=1,\ldots, r\}.
\]
If we do not care about computational complexity,
the dimension of $\calL_x$ 
can be calculated using the following simple formula:
\begin{eqnarray}
\label{inc-exc}
\dim{\calL_x} &=& \rk{I} - \sum_a \rk{H_a} +
\sum_{a<b} \rk{H_aH_b} -  \sum_{a<b<c} \rk{H_a H_b H_c} \nonumber \\
&& {} + \cdots
+ (-1)^r \rk{\prod_{a=1}^r H_a},
\end{eqnarray}
where $\rk{A}\equiv \dim{\Im{A}}$ is a rank of the operator $A$.
All summation here are carried out in the range $[1,r]$.
Formula~(\ref{inc-exc}) is analogous to
exclusion-inclusion formula for cardinality of a union of sets.
We can apply it since all projectors $H_a$ are diagonalizable
over the same basis and each projector can be identified with
the set of basis vectors which belong to $\Im{H_a}$.

Let $\Omega\subseteq\{1,\ldots,r\}$ be an arbitrary subset of check operators.
Denote
\begin{equation}
\label{r}
r(\Omega)=\rk{\prod_{a\in \Omega} H_a}.
\end{equation}
Formula~(\ref{inc-exc}) has the following important consequence.
Let $x=\{H_{a,j}\}$ and $x'=\{H_{a,j}'\}$ be two instances of the factorized
projectors \CES{} with the same $n$ and $r$. If for any subset
of check operators $\Omega$ the quantities $r(\Omega)$ for the instances $x$
and $x'$ coincide then both instances have the same answer.
So we can try to simplify the original instance $x$ by modifying the
projectors $H_{a,j}$ in such a way that all quantities $r(\Omega)$
are preserved. Although this approach seems to fail in a general
case (see a discussion at the end of this section), it works
perfectly for qubits.

In a case of qubits we have $\calH_j=\CC^2$ for all
$j$ and $\calH=(\CC^2)^{\otimes n}$.
Each operator $H_{a,j}\in \LinOp(\CC^2)$ is either the identity operator
or a projector of rank one. Let us fix the number of qubits $n$
and the number of check operators $r$.
Recall, that
the input $x=\{H_{a,j}\}$ is regarded as a table,
such that the columns correspond to the qubits and the rows correspond to
the check operators. We start from introducing an appropriate terminology.
\begin{dfn}
\label{def_commutative}
\rm
A table $x=\{H_{a,j}\}$ is called {\it commutative}
if $[H_a,H_b]=0$ for all $a$ and $b$.
\end{dfn}
\begin{dfn}
\label{def_consistent}
\rm
A table $x'=\{H_{a,j}'\}$ is called {\it consistent with}
a table $x=\{H_{a,j}\}$ if for any column $j$ one has
\begin{itemize}
\item $\rk{H_{a,j}}=\rk{H_{a,j}'}$ for all $a$.

\item $H_{a,j}=H_{b,j} \; \Rightarrow \; H_{a,j}'=H_{b,j}'$.

\item  $H_{a,j}H_{b,j}=0 \; \Rightarrow \; H_{a,j}'H_{b,j}'=0$.
\end{itemize}
\end{dfn}
\noindent
Two following lemmas show that we can substitute the original
table $x$ by any table $x'$ consistent with $x$ without
changing the answer of the problem.
\begin{lemma}
\label{lemma_commutative}
Let $x$ be a commutative table. If a table $x'$ is consistent
with $x$ then $x'$ is also a commutative table.
\end{lemma}
{\bf Proof:}
Let $x=\{H_{a,j}\}$, $x'=\{H_{a,j}'\}$,
$H_a=\bigotimes_{j=1}^n H_{a,j}$, and $H_a'=\bigotimes_{j=1}^n H_{a,j}'$.

Suppose  that $H_a$ and $H_b$ commute in a singular way i.e., $H_aH_b=0$.
It means that $H_{a,j}H_{b,j}=0$ for some $j$.
Since $x'$ is consistent with $x$, we have $H_{a,j}'H_{b,j}'=0$.
Thus $H_a'$ and $H_b'$ also commute (in a singular way).

Suppose now that $H_a$ and $H_b$ commute in a regular way, that is
$H_aH_b\ne 0$, $H_a H_b=H_b H_a$. It follows from Lemma~\ref{lemma_ab=ba}
that $H_{a,j}H_{b,j}=\pm H_{b,j}H_{a,j}$ for all $j$. Since 
both $H_{a,j}$ and $H_{b,j}$ are projectors, they can not
anticommute, so we conclude that $[H_{a,j},H_{b,j}]=0$ for all $j$.
Besides, we know that $H_{a,j}H_{b,j}\ne 0$. 
It is easy to see that 
both conditions can be met by one-qubit projectors only if 
for any fixed $j$ at least one of the following statements is true:

\noindent
(i) At least one of  $H_{a,j}$ and $H_{b,j}$ is the identity operator.

\noindent
(ii)  $H_{a,j}=H_{b,j}$.

\noindent
Now we can make use of the fact that $x'$ is consistent with $x$.
If the statement~(i) is true, one has $\rk{H_{a,j}}=2$ or (and) $\rk{H_{b,j}}=2$.
It follows that $\rk{H_{a,j}'}=2$ or (and) $\rk{H_{b,j}'}=2$, that is
at least one of the projectors $H_{a,j}'$  and $H_{b,j}'$ is the identity.
If the statement~(ii) is true, one has $H_{a,j}'=H_{b,j}'$.
In both cases 
$H_{a,j}'H_{b,j}'\ne 0$ and $[H_{a,j}',H_{b,j}']=0$. Since it holds for
all $j$, we conclude that
$H_a'$ and $H_b'$ commute (in a regular way).
\Square

\begin{lemma}
\label{lemma_consistent}
Let $x$ be a commutative table. If a table $x'$ is consistent with $x$
then all quantities $r(\Omega)$ for the tables $x$ and $x'$ coincide.
\end{lemma}
{\bf Proof:}
Let $x=\{H_{a,j}\}$, $x'=\{H_{a,j}'\}$,
$H_a=\bigotimes_{j=1}^n H_{a,j}$, and $H_a'=\bigotimes_{j=1}^n H_{a,j}'$.
According to Lemma~\ref{lemma_commutative} the table $x'$ is commutative,
so for any $\Omega$ we can define a quantity
\begin{equation}
r'(\Omega)=\rk{\prod_{a\in \Omega} H_a'}.
\end{equation}
We should prove that $r(\Omega)=r'(\Omega)$ for all $\Omega\subseteq\{1,\ldots,r\}$.
There are two possibilities:

\noindent
(i) $r(\Omega)>0$. It means that $H_a H_b\ne 0$ for all $a,b\in \Omega$.
Thus all operators $H_a$, $a\in \Omega$ commute in
a regular way and $[H_{a,j},H_{b,j}]=0$ for all $a,b\in \Omega$ and for all $j$.
In this situation the formula~(\ref{r}) for $r(\Omega)$ factorizes:
\begin {equation}
\label{r_factor}
r(\Omega)=\prod_{j=1}^n r_j(\Omega),\quad r_j(\Omega)= \rk{\prod_{a\in \Omega} H_{a,j}}.
\end{equation}
Let us consider some particular $j$. The family of projectors
$\{H_{a,j}\}_{a\in \Omega}$ is diagonalizable over the same basis.
Denote corresponding basis vectors as $|\psi_0\ra$ and $|\psi_1\ra$,
$\la \psi_\alpha|\psi_\beta\ra=\delta_{\alpha,\beta}$.
Each member of the family $\{H_{a,j}\}_{a\in \Omega}$
is one of the following projectors: $I$, $|\psi_0\ra\la \psi_0|$, and
$|\psi_1\ra\la\psi_1|$. The requirement $r_j(\Omega)>0$ implies that
the projectors $|\psi_0\ra\la \psi_0|$ and $|\psi_1\ra\la\psi_1|$
do not enter into this family simultaneously. Thus there
exist integers $k_1$ and $k_2$, $k_1+k_2=|\Omega|$, such that the  family
$\{H_{a,j}\}_{a\in \Omega}$ consists of $k_2$ identity operators $I$
and $k_1$ projectors of rank one $|\psi\ra\la \psi|$ (with
$|\psi\ra=|\psi_0\ra$ or $|\psi\ra=|\psi_1\ra$).
Now let us look at the family $\{H_{a,j}'\}_{a\in \Omega}$.
Since $x'$ is consistent with $x$, this family also consists of
$k_2$ identity operators $I$ and $k_1$ projectors of rank one
$|\varphi\ra\la \varphi|$ for some $|\varphi\ra\in \CC^2$.
Therefore $[H_{a,j}',H_{b,j}']=0$ for all $a,b\in \Omega$ and
$$
r_j'(\Omega)=\rk{\prod_{a\in \Omega} H_{a,j}'}=r_j(\Omega).
$$
Also it means that the quantity $r'(\Omega)$ factorizes,
$r'(\Omega)=\prod_{j=1}^n r_j'(\Omega)$, and thus $r'(\Omega)=r(\Omega)$.

\noindent
(ii) $r(\Omega)=0$. It means that $\prod_{a\in \Omega} H_a=0$.
Suppose first that $H_a H_b=0$ for some $a,b\in \Omega$.
Since $x'$ is consistent with $x$ it implies that $H_a' H_b'=0$
(see the last part of the proof of Lemma~\ref{lemma_commutative})
and so that $r'(\Omega)=0$. Now suppose that $H_a H_b\ne 0$ for
all $a,b\in \Omega$. By definition, it means that all
check operators $H_a$, $a\in \Omega$ commute in a regular way,
i.e. $[H_{a,j},H_{b,j}]=0$ for all $a,b\in \Omega$ and for all $j$.
In particular, the family $\{H_{a,j}\}_{a\in \Omega}$ is
diagonalizable over the same basis.
In this situation we can use a decomposition~(\ref{r_factor}).
We know that $r_j(\Omega)=0$ for some $j$. But it happens iff
the family $\{H_{a,j}\}_{a\in \Omega}$ contains a pair
of rank one projectors corresponding to  mutually orthogonal states,
i.e. $H_{a,j}H_{b,j}=0$ for some $a,b\in \Omega$.
But it implies $H_aH_b=0$ which contradicts our assumption.
\Square

What is the most simple form of a table $x'$ consistent with the
original table $x$? We will show that for any table
$x$ (which may be not a commutative one) there exists
a table $x'=\{H_{a,j}'\}$ consistent with $x$ such that
$H_{a,j}'\in \{I, |0\ra\la 0|,|1\ra\la 1|\}$ for all $a$ and $j$.
Here $|0\ra,|1\ra\in \CC^2$ is some fixed orthonormal basis of $\CC^2$
(computational basis).
All check operators $H_a'$ for the table $x'$ are diagonal in the
computational basis of $(\CC^2)^{\otimes n}$, therefore Merlin's proof
might be a description of the table $x'$ and
a binary string $(x_1,x_2,\ldots,x_n)$
such that $H_a'|x_1\ra\otimes|x_2\ra\otimes\cdots\otimes|x_n\ra=0$ for all $a$.
Verification that $x'$ is indeed consistent with $x$
requires only $O(nr^2)$ computational steps. Thus existence of
a table $x'$ with the specified properties implies that the
factorized projectors \CES{} for qubits belongs to \NP.
It remains to prove the following lemma.

\begin{lemma}
\label{lemma_graph}
For any table $x$ there exists a table $x'=\{H_{a,j}'\}$
consistent with $x$ such that $H_{a,j}'\in \{I, |0\ra\la 0|,|1\ra\la 1|\}$
for all $a$ and $j$.
\end{lemma}
{\bf Proof:}
Let $x=\{H_{a,j}\}$.
A transformation from $x$ to the desired table $x'$ is defined independently
for each column, so let us focus on some particular
column, say $j=1$. At first, we define an {\it orthogonality graph} $G=(V,E)$.
A vertex $v\in V$ is a set of rows which contain the same
projector. In other words, we introduce an equivalence relation
on the set of rows: $a\sim b\; \Leftrightarrow \; H_{a,1}=H_{b,1}$
and define a vertex $v\in V$ as an equivalence class of rows.
Thus, by definition, each vertex $v\in V$ carries a projector $H(v)\in \LinOp(\CC^2)$.
A pair of vertices $u,v\in V$ is connected by an edge iff the projectors
corresponding to $u$ and $v$ are orthogonal:
$(u,v)\in E \; \Leftrightarrow \; H(u)H(v)=0$.

Consider as an example the following table ($r=100$): $H_{1,1}=I$,
$H_{2,1}=H_{3,1}=1/2(I+\sz)$,
$H_{4,1}=1/2(I-\sz)$,
$H_{5,1}=1/2(I+\sx)$,
$H_{6,1}=1/2(I-\sx)$,
$H_{7,1}=\cdots=H_{100,1}=1/2(I+\sy)$.
Then an orthogonality graph consists of six vertices, $V=\{1,2,3,4,5,6\}$,
with $H(1)=I$, $H(2)=1/2(I+\sz)$,
$H(3)=1/2(I-\sz)$,
$H(4)=1/2(I+\sx)$,
$H(5)=1/2(I-\sx)$, and
$H(6)=1/2(I+\sy)$.
The set of edges is $E=\{(2,3),(4,5)\}$.

It is a special property of qubits that any orthogonality graph
always splits to several disconnected edges representing pairs of orthogonal
projectors and several disconnected vertices representing
unpaired projectors of rank one and the identity operator.

Suppose we perform a transformation
\begin {equation}
\label{transform1}
H(v)\to H'(v), \quad v\in V,
\end{equation}
for some projectors $H'(v)\in \LinOp(\CC^2)$ which satisfy
\begin {equation}
\label{transform2}
\rk{H(v)}=\rk{H'(v)}\quad \mbox{for all } v\in V;
\quad H'(u)H'(v)=0 \quad \mbox{for all } (u,v)\in E.
\end{equation}
As each vertex of the graph represents a group of cells of
the table, the transformation~(\ref{transform1}) can be also regarded
as a transformation of the tables $x\to x'$.
Note that the table $x'$ is consistent with the table $x$, since the
restrictions~(\ref{transform2}) are just rephrasing of Definition~\ref{def_consistent}.

Now existence of the table $x'$ with the desired properties is obvious.
For each disconnected edge $(u,v)\in E$ we define the transformation~(\ref{transform1})
as $H'(u)=|0\ra\la 0|$, $H'(v)=|1\ra\la1|$ (it does not matter, how exactly
$0$ and $1$ are assigned to endpoints of the edge).
For any disconnected vertex $v\in V$, we define $H'(v)=I$ if $H(v)=I$
and $H'(v)=|0\ra\la 0|$ if $\rk{H(v)}=1$.
\Square

We conclude this section by several remarks concerning the factorized projectors
\CES{} problem with $d>2$. For simplicity, let us put an additional constraint,
namely
that each projector $H_{a,j}$ is either the identity operators or a projector of
rank one (a projector on a pure state). Definitions~\ref{def_commutative}
and~\ref{def_consistent} are still reasonable in this setting. Moreover,
it is easy to check that Lemmas~\ref{lemma_commutative} and~\ref{lemma_consistent}
are still valid (the proofs given above can be repeated almost literally).
A natural generalization of Lemma~\ref{lemma_graph} might be the following:

\noindent {\it
For any table $x$ there exists a table
$x'=\{H_{a,j}'\}$
consistent with $x$ such that for all $a$ and $j$
 $H_{a,j}'\in \{I, |1\ra\la 1|,\ldots,|d\ra\la d|\}$.}

\noindent
Here some fixed orthonormal basis $|1\ra,\ldots,|d\ra\in \CC^d$ is
chosen. Unfortunately, this statement is wrong even for $d=3$.
Counterexamples may be obtained by constructions used in the proof
of the Kochen-Specker theorem~\cite{KS}. According to this theorem
there exist families of projectors $P_1,\ldots,P_r \in \LinOp(\CC^d)$ ($d\ge 3$)
which do not admit an assignment
\begin{equation}
\label{Kochen_Specker1}
P_a \to \eps_a \in \{0,1 \}, \quad a=1,\ldots,r,
\end{equation}
such that
\begin{equation}
\label{Kochen_Specker2}
\sum_{a\in \Omega} \eps_a =1 \quad \mbox{whenever }
\sum_{a\in \Omega} P_a = I.
\end{equation}
Here $\Omega\subseteq\{1,\ldots,r\}$ may be an arbitrary subset.
Peres~\cite{Peres} suggested an explicit construction of such family for
$d=3$ and $r=33$. This family consists of the projectors of rank one,
i.e. $P_a=|\psi_a\ra\la\psi_a|$, $|\psi_a\ra\in \CC^3$, $a=1,\ldots,33$.

Suppose a table $x=\{H_{a,j}\}$ consists of $33$ rows and the first column
accommodates the family of projectors  suggested by Peres:
$H_{a,1}=|\psi_a\ra\la\psi_a|$, $a=1,\ldots,33$. Let $x'=\{H_{a,j}'\}$ be a table
whose existence is promised by the generalized Lemma~\ref{lemma_graph}.
Since $x'$ is consistent with $x$, one has $\rk{H_{a,1}'}=\rk{H_{a,1}}=1$,
so neither of the projectors $H_{a,1}'$, $a=1,\ldots,33$, is the identity.
Then the only possibility (if the lemma is true) is that
$H_{a,1}'\in \{ |1\ra\la 1|, |2\ra\la 2|, |3\ra\la 3|\}$.
A  consistency property implies also that
\begin{equation}
\label{Kochen_Specker3}
\sum_{a\in \Omega} H_{a,1} = I \quad \Rightarrow \quad
\sum_{a\in \Omega} H_{a,1}' = I.
\end{equation}
Indeed, the equality on the lefthand side is possible iff $|\Omega|=3$ and
all projectors $\{H_{a,1}\}_{a\in \Omega}$ are pairwise orthogonal.
Then the projectors $\{H_{a,1}'\}_{a\in \Omega}$ are also pairwise orthogonal
and we get the equality on the righthand side.
The family of projectors $\{H_{a,1}'\}$ obviously admits an
assignment~(\ref{Kochen_Specker1},\ref{Kochen_Specker2}).
Indeed, we can put
$$
\eps_a = \left\{ \ba{rcl}
1 &\mbox{if}& H_{a,1}'=|3\ra\la 3|,\\
0 &\mbox{if}& H_{a,1}'=|1\ra\la 1| \; \mbox{or} \; |2\ra\la 2|.\\
\ea \right.
$$
But the property~(\ref{Kochen_Specker3}) implies that the assignment
$H_{a,1}\to \eps_a$, $a=1,\ldots,33$ also satisfies the requirements~(\ref{Kochen_Specker2}).
It is impossible. Therefore the generalization of Lemma~\ref{lemma_graph} given above
is wrong.

In fact, the proof of Lemma~\ref{lemma_graph} needs a regular
$d$-coloring of a graph which admits $d$-dimensional orthogonal representation.
As we have seen, this is not always possible.  It might happen however that
all `pathological' (which violate Lemma~\ref{lemma_graph})
{\it commutative} tables lead to simple instances
of factorized projectors \CES. Indeed, a difficult instance
must contain pairs of rows commuting
in a singular way and pairs commuting in a regular way.
The number of pairs of each type must be sufficiently large.
For example, if all rows
commute in a regular way, the problem belongs to \NP{}
according to Corollary~\ref{corrol1}.
If all rows commute in a singular way, we can easy compute
$\dim{\calL_0}$ using the exclusion-inclusion formula~(\ref{inc-exc}).
The number of `pathological' columns in the table also must be sufficiently large.
To construct difficult instances we must meet all these
requirements which seems to be hard.

\nonumsection{Acknowledgements}
\noindent

We would like to thank P. Wocjan for interesting discussions which
motivated this line of research. We thank A. Kitaev and J. Preskill for helpful
comments and suggestions.
We are grateful to the referee of the paper for numerous remarks and
corrections.
The main part of this work was done when M.V. was visiting
Institute for Quantum Information, Caltech. The work was  supported by RFBR
grant 02-01-00547, and
by the National Science Foundation under Grant No.\ EIA-0086038.

\nonumsection{References}


\begin{thebibliography}{000}
\bibitem{referee} This Lemma and its proof
were suggested by the referee of the paper.
\end{thebibliography}
\end{document}